\documentclass[12pt]{article}
\pdfoutput=1
\usepackage[a4paper]{geometry}

\usepackage{jheppub, amsmath,amssymb,amsfonts,amsxtra, mathrsfs, makeidx,graphics,graphicx,amsthm,epsfig, bm,longtable,float, color,tikz,mathtools,xfrac,footnote,rotating, lscape, makecell, environ,mathtools, empheq}

\usepackage{enumitem}

\usepackage{subfig}
\usepackage{multirow}
\usepackage{adjustbox}

\usepackage{hyperref}

\usepackage{commath}
\usepackage[english]{babel}
\usepackage{setspace}

\usepackage{amscd,color}

\usepackage{verbatim,booktabs}

\usepackage{latexsym}

\usepackage{youngtab}

\usepackage{pdflscape}

\DeclareMathOperator{\Tr}{Tr}
\DeclareMathOperator{\rk}{rk}

\newcommand{\zz}{\mathbb{Z}}
\newcommand{\nn}{\mathcal{N}}
\newcommand{\nc}{N_c}

\usepackage{tikz}
\usetikzlibrary{matrix}
\usetikzlibrary{tikzmark}
\usetikzlibrary{math} 
\usetikzlibrary{calc}
\usepackage{xifthen}

\definecolor{mandarancio}{RGB}{255,147,0}
\definecolor{azzurro}{RGB}{0,118,186}
\definecolor{terrabruciatadiconcorezzo}{RGB}{181,23,0}

\def\centerarc[#1](#2)(#3:#4:#5)
{ \draw[#1] ($(#2)+({#5*cos(#3)},{#5*sin(#3)})$) arc (#3:#4:#5) }

\newcommand{\LeftRightQuiver}[4]{
	\begin{tikzpicture}[anchor=center,baseline=-0.5ex, scale=1.55, thick]
		\tikzmath{
			\sep = 0.7; \nradius=0.2; \tradius = 0.2; \doublelinedelta=0.02;
			\tlcenter = -\nradius -\tradius+ 0.08; \trcenter =\sep +\nradius +\tradius- 0.08;
			\labheight = \nradius + 0.15;
			\leftlabel = -\nradius -\tradius - 0.12;	\rightlabel =\sep - (-\nradius -\tradius - 0.12); 
			\lclip = -\nradius -\tradius - 0.65;			\rclip = \sep - (-\nradius -\tradius - 0.7); 
			\tradiusext = \tradius + \doublelinedelta; \tradiusint = \tradius - \doublelinedelta;	}
		\clip (\lclip,-\nradius*2) rectangle (\rclip,\labheight + 0.2);
		\draw (0,0)  circle [radius=\nradius];
		\node at (0,\labheight) {#1};
		\draw (\sep,0)  circle [radius=\nradius];
		\node at (\sep,\labheight) {#2};
		\draw[dashed] (\nradius,0) -- (\sep - \nradius,0);
		\ifthenelse{\equal{#3}{$\phi$}}                    
		{\centerarc[<-](\tlcenter,0)(360-45:45:\tradius);}{}
		\ifthenelse{\equal{#3}{$S$}}                    
		{\centerarc[orange](\tlcenter,0)(360-45:45:\tradius);}{}
		\ifthenelse{\equal{#3}{$A$}}                     
		{\centerarc[azzurro](\tlcenter,0)(360-45:45:\tradius);} {}
		\ifthenelse{\equal{#3}{$S,\tilde{S}$}}                     
		{\centerarc[orange](\tlcenter,0)(360-45:45:\tradiusext);
			\centerarc[orange,dashed](\tlcenter,0)(360-45:45:\tradiusint);}{}
		\ifthenelse{\equal{#3}{$A,\tilde{A}$}}                     
		{\centerarc[azzurro](\tlcenter,0)(360-45:45:\tradiusext);
			\centerarc[azzurro,dashed](\tlcenter,0)(360-45:45:\tradiusint);}{}
		\node[anchor=east] at (\leftlabel,0) {#3};
		\ifthenelse{\equal{#4}{$\phi$}}                    
		{\centerarc[<-](\trcenter,0)(-135:135:\tradius);}{}
		\ifthenelse{\equal{#4}{$S$}}                    
		{\centerarc[orange](\trcenter,0)(-135:135:\tradius);}{}
		\ifthenelse{\equal{#4}{$A$}}                     
		{\centerarc[azzurro](\trcenter,0)(-135:135:\tradius);} {}
		\ifthenelse{\equal{#4}{$S,\tilde{S}$}}                     
		{\centerarc[orange](\trcenter,0)(360-45:45:\tradiusext);
			\centerarc[orange,dashed](\trcenter,0)(-135:135:\tradiusint);}{}
		\ifthenelse{\equal{#4}{$A,\tilde{A}$}}                     
		{\centerarc[azzurro](\trcenter,0)(-135:135:\tradiusext);
			\centerarc[azzurro,dashed](\trcenter,0)(-135:135:\tradiusint);}{}
		\node[anchor=west] at (\rightlabel,0) {#4};
	\end{tikzpicture}
}

\newcommand{\SimpElliptic}[2]{ 
	\begin{tikzpicture}[anchor=center,baseline=-0.5ex]
		\clip (-1.85,-0.5) rectangle (1.85,0.5);
		\draw[azzurro] (0,0) ellipse (1.5 and 0.4);
		\draw[mandarancio] (-1.8,-0.5) -- (-1.8,.4) -- (-1.2,.5) -- (-1.2,-.4) -- (-1.8,-0.5);
		\draw[mandarancio] (1.8,-0.4) -- (1.8,.5) -- (1.2,.4) -- (1.2,-.5) -- (1.8,-0.4);
		\ifthenelse{\isempty{#1}}%
		{}
		{\draw[terrabruciatadiconcorezzo, thick] (-1.5,-0.45) -- (-1.5,0.45); }
		\ifthenelse{\isempty{#2}}%
		{}
		{\draw[terrabruciatadiconcorezzo, thick] (1.5,-0.45) -- (1.5,0.45); }
		\begin{scope}
			\pgfset{minimum size=1.5cm,inner sep=2mm}
			\pgfnode{rectangle}{center}{}{nodename}{\pgfusepath{clip}}
			\fill[white] (-1.5cm,-.5cm) rectangle (1.5cm, .5cm);
			\draw[azzurro, dashed] (0,0) ellipse (1.5 and 0.4);
		\end{scope}
	\end{tikzpicture}
}

\newcommand{\Quiver}[7]{
	\begin{tikzpicture}[scale=#6, thick]
		\tikzmath{
			\rnode = 0.5;	\sep = 3.5; 	\dotsep = 0.5;	\fundsep = 3pt;	\rtens = 0.5;	 \doublelinedelta=0.04;
			\crtens = 3*\sep + \dotsep + \rtens + \rnode - 0.15;	\trlabel = \crtens + \rtens + 0.1;
			\cltens = -\rtens - \rnode + 0.15;	\tllabel = \cltens - \rtens - 0.1;
			\rnodeext = \rnode + 0.05;
			\labelheight=\rnode + 0.1;
			\rtensext = \rtens + \doublelinedelta;	\rtensint = \rtens - \doublelinedelta;
		}
		\draw (0,0) circle [radius=\rnode];
		\draw[yshift=\fundsep, ->] (\rnodeext,0) -- (\sep-\rnodeext,0);
		\draw[yshift=-\fundsep, <-] (\rnodeext,0) -- (\sep-\rnodeext,0);
		\draw (\sep,0) circle [radius=\rnode];
		\begin{scope}[xshift=\sep cm]
			\draw[yshift=\fundsep, ->] (\rnodeext,0) -- (\sep-\rnodeext,0);
			\draw[yshift=-\fundsep, <-] (\rnodeext,0) -- (\sep-\rnodeext,0);
		\end{scope}
		\draw (2*\sep,0) circle [radius=\rnode];
		
		\draw[dashed] (2*\sep+\rnode,0) -- (3*\sep + \dotsep-\rnode,0);
		\draw (3*\sep + \dotsep,0) circle [radius=\rnode];
		
		\draw (0,\labelheight) 				node[anchor=south]{#1};
		\draw (\sep,\labelheight) 			node[anchor=south]{#2};
		\draw (2*\sep,\labelheight) 		node[anchor=south]{#3};
		\draw (3*\sep+\dotsep,\labelheight) node[anchor=south]{#4};
		
		\ifthenelse{\equal{#5}{$A,\tilde{A}$}}%
		{
			\centerarc[azzurro](\crtens,0)(-135:135:\rtensext);
			\centerarc[azzurro, dashed](\crtens,0)(-135:135:\rtensint);
			\node[anchor=west] at (\trlabel,0) {#5};} {}
		\ifthenelse{\equal{#5}{$S,\tilde{S}$}}%
		{
			\centerarc[orange](\crtens,0)(-135:135:\rtensext);
			\centerarc[orange, dashed](\crtens,0)(-135:135:\rtensint);
			\node[anchor=west] at (\trlabel,0) {#5};} {}
		
		\ifthenelse{\equal{#7}{$A,\tilde{A}$}}%
		{
			\centerarc[azzurro](\cltens,0)(360-45:45:\rtensext);
			\centerarc[azzurro, dashed](\cltens,0)(360-45:455:\rtensint);
			\node[anchor=east] at (\tllabel,0) {#7};} {}
		\ifthenelse{\equal{#7}{$S,\tilde{S}$}}%
		{
			\centerarc[orange](\cltens,0)(360-45:45:\rtensext);
			\centerarc[orange, dashed](\cltens,0)(360-45:45:\rtensint);
			\node[anchor=east] at (\tllabel,0) {#7};} {}
		
	\end{tikzpicture}
}

\newcommand{\QuiverNtwo}[7]{
	\begin{tikzpicture}[scale=#6, thick]
		\tikzmath{
			\rnode = 0.5;	\sep = 3.5; 	\dotsep = 0.5;	\fundsep = 3pt;	\rtens = 0.5;	 \doublelinedelta=0.04;
			\crtens = 3*\sep + \dotsep + \rtens + \rnode - 0.15;	\trlabel = \crtens + \rtens + 0.1;
			\cltens = -\rtens - \rnode + 0.15;	\tllabel = \cltens - \rtens - 0.1;
			\adjheight = -\rtens - \rnode + 0.15;
			\rnodeext = \rnode + 0.05;
			\labelheight=\rnode + 0.1;
			\rtensext = \rtens + \doublelinedelta;	\rtensint = \rtens - \doublelinedelta;
		}
		\draw (0,0) circle [radius=\rnode];
		\draw[yshift=\fundsep, ->] (\rnodeext,0) -- (\sep-\rnodeext,0);
		\draw[yshift=-\fundsep, <-] (\rnodeext,0) -- (\sep-\rnodeext,0);
		\draw (\sep,0) circle [radius=\rnode];
		\begin{scope}[xshift=\sep cm]
			\draw[yshift=\fundsep, ->] (\rnodeext,0) -- (\sep-\rnodeext,0);
			\draw[yshift=-\fundsep, <-] (\rnodeext,0) -- (\sep-\rnodeext,0);
		\end{scope}
		\draw (2*\sep,0) circle [radius=\rnode];
		
		\draw[dashed] (2*\sep+\rnode,0) -- (3*\sep + \dotsep-\rnode,0);
		\draw (3*\sep + \dotsep,0) circle [radius=\rnode];
		
		\draw (0,\labelheight) 				node[anchor=south]{#1};
		\draw (\sep,\labelheight) 			node[anchor=south]{#2};
		\draw (2*\sep,\labelheight) 		node[anchor=south]{#3};
		\draw (3*\sep+\dotsep,\labelheight) node[anchor=south]{#4};
		
		\ifthenelse{\equal{#5}{$A,\tilde{A}$}}%
		{
			\centerarc[azzurro](\crtens,0)(-135:135:\rtensext);
			\centerarc[azzurro, dashed](\crtens,0)(-135:135:\rtensint);
			\node[anchor=west] at (\trlabel,0) {#5};} {}
		\ifthenelse{\equal{#5}{$S,\tilde{S}$}}%
		{
			\centerarc[orange](\crtens,0)(-135:135:\rtensext);
			\centerarc[orange, dashed](\crtens,0)(-135:135:\rtensint);
			\node[anchor=west] at (\trlabel,0) {#5};} {}
		
		\ifthenelse{\equal{#7}{$A,\tilde{A}$}}%
		{
			\centerarc[azzurro](\cltens,0)(360-45:45:\rtensext);
			\centerarc[azzurro, dashed](\cltens,0)(360-45:455:\rtensint);
			\node[anchor=east] at (\tllabel,0) {#7};} {}
		\ifthenelse{\equal{#7}{$S,\tilde{S}$}}%
		{
			\centerarc[orange](\cltens,0)(360-45:45:\rtensext);
			\centerarc[orange, dashed](\cltens,0)(360-45:45:\rtensint);
			\node[anchor=east] at (\tllabel,0) {#7};} {}
		
		\centerarc[->](0,\adjheight)(-180-45:45:\rtens);
		\centerarc[->](\sep,\adjheight)(-180-45:45:\rtens);
		\centerarc[->](2*\sep,\adjheight)(-180-45:45:\rtens);
		\centerarc[->](3*\sep+\dotsep,\adjheight)(-180-45:45:\rtens);
		
		\node[anchor=west] at (\rtens,\adjheight-0.1) {$\phi_1$};
		\node[anchor=west] at (\sep+\rtens,\adjheight-0.1) {$\phi_2$};	
		\node[anchor=west] at (2*\sep+\rtens,\adjheight-0.1) {$\phi_3$};	
		\node[anchor=west] at (3*\sep+\dotsep+\rtens,\adjheight-0.1) {$\phi_{n_g+1}$};
	\end{tikzpicture}
}

\title{
\begin{center}
Conformal S-dualities from O-planes
\end{center}
}

\author[a]{Antonio Amariti,}
\author[b,c]{Marco Fazzi,} 
\author[a,d]{Simone Rota,}
\author[a,d]{and Alessia Segati}

\affiliation[a]{INFN, Sezione di Milano, Via Celoria 16, I-20133 Milano, Italy}
\affiliation[b]{INFN, Sezione di Milano-Bicocca, Piazza della Scienza 3, I-20126 Milano, Italy}
\affiliation[c]{Dipartimento di Fisica, Universit\`a di Milano-Bicocca, Piazza della Scienza 3, I-20126 Milano, Italy}
\affiliation[d]{Dipartimento di Fisica, Universit\`a degli Studi di Milano, Via Celoria 16, I-20133 Milano, Italy}

\emailAdd{antonio.amariti@mi.infn.it,  marco.fazzi@mib.infn.it,  simone.rota@mi.infn.it, alessia.segati@mi.infn.it}

\abstract{ 
We study 4d SCFTs obtained by orientifold projections on necklace quivers with fractional branes. The models obtained by this procedure are $\mathcal{N}=1$ linear quivers with unitary, symplectic and orthogonal  gauge groups, bifundamental and tensorial matter.  Remarkably, models that are not dual in the unoriented case can have the same central charges and superconformal index after the projection. The reason for this behavior rests upon the ubiquitous presence of adjoint fields with R-charge one.  We claim that the presence of such fields is at the origin of the notion of inherited S-duality on the models' conformal manifold.
}

\begin{document}

\maketitle

\section{Introduction}
\label{sec:intro}

The original S-duality conjecture, due to Montonen and Olive \cite{Montonen:1977sn}, states that $\mathcal{N}=4$ SYM 
with gauge group $G$ is equivalently described by the Langlands S-dual gauge group $^L G$
provided that the exactly marginal holomorphic gauge coupling $\tau$ is opportunely transformed into $-1/\tau$.
More generally the group of transformations of $\tau$ is enlarged to $SL(2,\mathbb{Z})$ (or its extension to the non-simply laced cases)
if the transformation $\tau \rightarrow \tau+1$ is taken into account. The full classification of all possible gauge groups
associated to these transformations of $\tau$ has been provided in \cite{Aharony:2013hda}, in terms  of charge lattices of mutually local Wilson and~'t Hooft line operators.

S-duality has then been extended to  4d models with eight supercharges (see for example \cite{Seiberg:1994aj,Donagi:1995cf,Argyres:1995wt,Argyres:2007cn,Gaiotto:2009we}). In this case  it has been either associated to 
a motion in the space of coupling constants (generalizing the $\mathcal{N}=4$ case) or to a motion in the moduli space (this is for example the case of the Seiberg--Witten theory). Here we will be interested in the first generalization.

A  generalization to $\mathcal{N}=2$ elliptic models has been provided in \cite{Donagi:1995cf}. These are necklace 
quiver gauge theories with $\mathcal{N}=2$ $SU(N_c)$ vector multiplets, with bifundamental hypermultiplets connecting consecutive nodes.
In this case the S-duality group corresponds to the mapping class group 
of the punctured torus that emerges when the model is described in M-theory.

$\mathcal{N}=4$ S-duality has also been generalized to $\mathcal{N}=1$ theories by an opportune mass deformation. This mechanism, denoted  inherited duality, has been extended in \cite{Halmagyi:2004ju} to elliptic models.
In this case it has been pointed out that the inherited S-duality for $\mathcal{N}=1$ elliptic models 
coincides with the $\mathcal{N}=2$ S-duality group. It was furthermore shown that it 
is in general distinct from Seiberg duality,  even if the latter, in some cases, corresponds to the
action of a subgroup of the inherited S-duality group.

Such a construction generalizes the fact that S-duality for $\mathcal{N}=2$ $SU(N_c)$ SQCD with $N_f=2N_c$ 
reduces to Seiberg duality after perturbing the superpotential with a mass term for the adjoint field in the $\mathcal{N}=2$ vector multiplet
 \cite{Leigh:1995ep,Strassler:2005qs}.

The mechanism of inherited duality can naturally be extended to other $\mathcal{N}=2$  models obtained from the elliptic models
when considering the action of orientifolds. For example, if pairs of O6-planes are considered in the Type IIA description of the
elliptic models, new S-dual pairs have been conjectured in \cite{Uranga:1998uj}.
In these cases it is also crucial to add fractional branes in order to guarantee the cancellation of gauge anomalies.

Similar orientifold projections can also be applied to the $\mathcal{N}=1$ version of the elliptic models. The latter are often referred to as the (infinite family of) $L^{pqp}$ models, because they describe the near-horizon limit of D3-branes probing the tip of a Calabi--Yau cone over a five-dimensional Sasaki--Einstein $L^{pqp}$ base (the $p=0$ case corresponding to the 
$\mathcal{N}=2$ elliptic models) \cite{Benvenuti:2005ja,Butti:2005sw,Franco:2005sm}.

In this fashion the orientifolds can be equivalently studied by using the dimer models techniques developed in \cite{Franco:2007ii}.
This analysis has been pursued in great detail in a recent paper \cite{Antinucci:2021edv}, where it has been observed that there are infinite families 
of examples in which models that are unrelated by any IR duality before the projection turn out to share the same central charges and superconformal indices in the IR after it. This generalizes a previous construction \cite{Antinucci:2020yki} for oriented chiral quivers.
The equivalence of these quantities led to a natural duality conjecture among these oriented models, but this duality  cannot be always obtained by applying the usual rules of Seiberg duality on the quiver.

The aim of this paper is to interpret the appearance of such dualities in terms of inherited S-duality from the $\mathcal{N}=2$ case of \cite{Uranga:1998uj}. Inspired by the case of $SU(N_c)$ Seiberg duality with $N_f=2N_c$, here  we observe that the duality for the $\mathcal{N}=1$ models discussed in \cite{Antinucci:2021edv}   corresponds to S-duality at different points of the conformal manifold. The expected inherited S-dual models can be obtained from the ones of \cite{Uranga:1998uj}
 by a motion in the conformal manifold. The crucial property behind this result is the presence in the spectrum of 
 two-index tensor fields with R-charge equal to one, uncharged with respect to the other global
symmetries \cite{Leigh:1995ep}. The presence of two-index tensors (mostly adjoints) with this property is 
ubiquitous in the models under investigation and it implies that one can always turn on quadratic deformations involving  such   fields.
These quadratic deformations are exactly marginal and they are involved in the parameterization of the conformal manifold.

A further check of this claim consists of integrating out the \textit{conformal} masses (even if the interpretation of this integration is not completely clear) and reformulate all the dual phases of \cite{Antinucci:2021edv} in a self-dual form, as expected by inheriting the S-duality from the construction of \cite{Uranga:1998uj}.
Furthermore, to corroborate the results, we study other inherited S-dualities that are obtained from the construction of \cite{Antinucci:2021edv}, where the models do not reduce to a self-dual form. This is the analog of S-duality between $B_n$ and $C_n$ $\mathcal{N}=4$ SYM theories.
We then check the validity of our results by matching the central charges, the 't Hooft anomalies, and in some small-rank cases
the superconformal indices by expanding them in the R-symmetry fugacities.

The paper is organized as follows. In section \ref{sec:remarks} we discuss some known aspects of inherited S-duality for $\mathcal{N}=1$ SCFTs
and then we review the case of $SU(N_c)$ SQCD with $N_f=2N_c$, defining the notion of S-duality on the conformal manifold in presence of fields with R-charge equal to one.
In section \ref{sec:elliptic} we review basic aspects of S-duality for necklace quivers in the absence of O-planes and discuss inherited S-duality in this case. We also review the construction of oriented $\mathcal{N}=2$ necklaces, by the addition of pairs of O6-planes, reviewing known facts about S-duality for such models.
Sections \ref{sec:newind} and \ref{sec:exam} are the core of the paper, where we construct the new $\mathcal{N}=1$ S-dualities for the oriented necklaces. We first construct the $\mathcal{N}=1$ models, then we propose the duality, and check them by the means discussed above. 
In section \ref{sec:conc} we conclude with a few remarks and discuss further lines of research.

%
%
%
\section{Remarks on inherited S-duality}
\label{sec:remarks}
%
%
%
In this section we briefly review the argument in favor of an $SL(2,\mathbb{Z})$ invariance of 
$\mathcal{N}=1$ SCFTs originating from S-duality.
The basic example proposed in \cite{Leigh:1995ep} and then proven in \cite{Argyres:1999xu} corresponds to an $\mathcal{N}=1$ 
gauge theory with gauge group $G$ and two adjoints $\Phi_1$ and $\Phi_2$  interacting through an exactly marginal superpotential 
$W = h [\Phi_1,\Phi_2]^2$, where traces are understood.
The claim of \cite{Leigh:1995ep} was that this theory exhibits an $SL(2,\mathbb{Z})$ invariance inherited from the one of 
$\mathcal{N}=4$ SYM modified by a massive superpotential, i.e. 
\begin{equation}
\label{intri}
W = \sqrt 2 \Phi_1 [\Phi_2,\Phi_3]+ \frac{m_3}{2} \Phi_3^2\ .
\end{equation}
In order to prove this claim in \cite{Argyres:1999xu} the authors added to the superpotential (\ref{intri}) another mass term, say for $\Phi_2$, 
and observed that the holomorphic quantities depend only on the product $m_2 m_3$.
This allowed them to identify the marginal parameter of the $\mathcal{N}=1$ gauge theory with the holomorphic gauge coupling of the $\mathcal{N}=4$ theory.
Thanks to this mechanism they showed that the $SL(2,\mathbb{Z})$ invariance was inherited by the $\mathcal{N}=1$ theory.

An analogous discussion has been pursued in \cite{Leigh:1995ep,Strassler:2005qs}  for the case of $\mathcal{N}=2$ $N_f=2N_c$ SQCD. 
In this case the inherited S-duality corresponds to Seiberg duality.
This last case will play a crucial role in our analysis and for this reason we will now discuss it in more detail.

\subsection{\texorpdfstring{A lesson from $\mathcal{N}=1$ SQCD}{A lesson from N=1 SQCD}}

\begin{figure}
\begin{center}
\includegraphics[width=10cm]{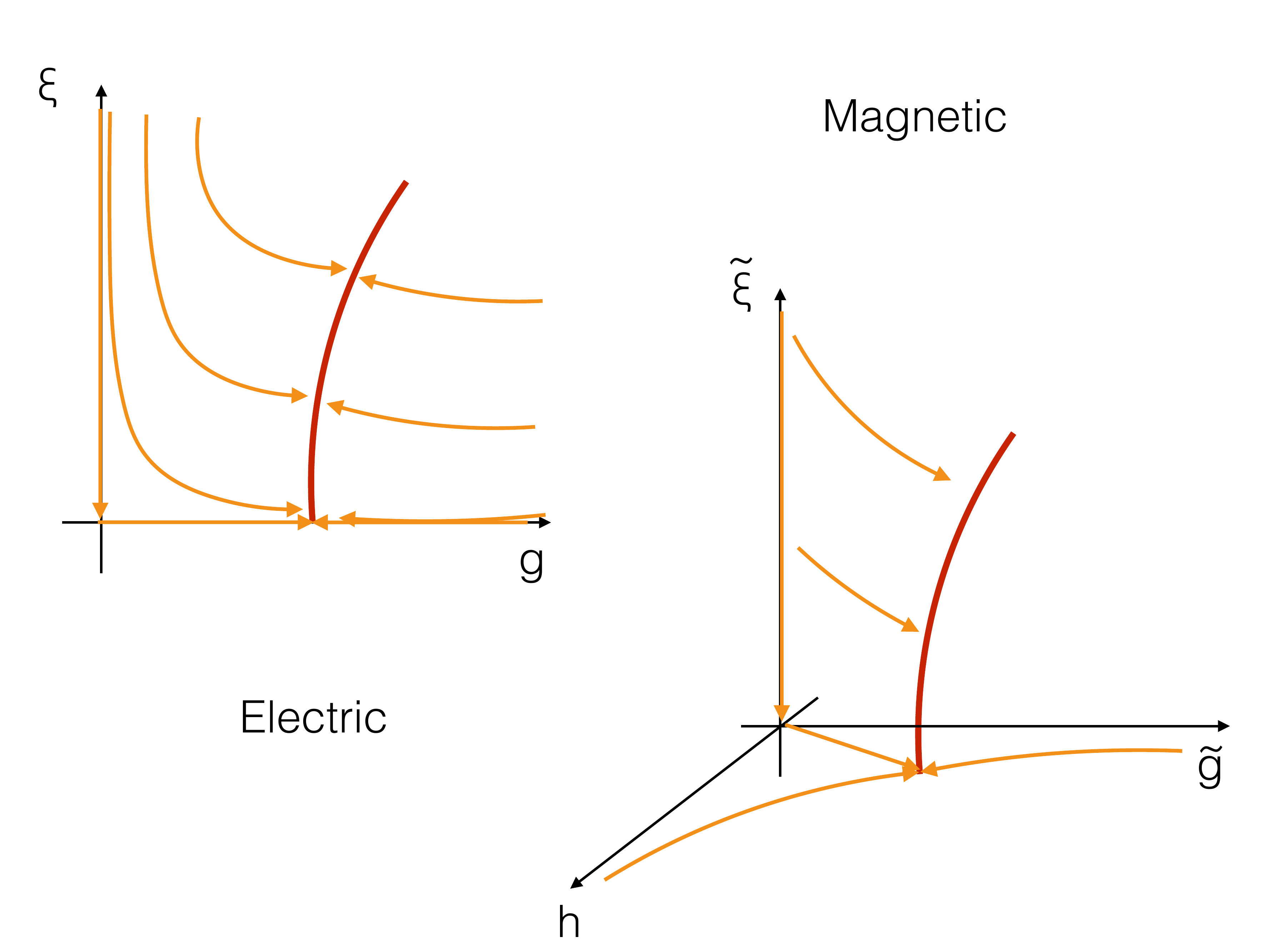}
\caption{On the LHS: couplings of the electric SQCD for $N_f=2N_c$ SQCD deformed by a quartic superpotential for the quarks. On the RHS couplings of the dual magnetic SQCD for $N_f=2N_c$ SQCD deformed by a quadratic superpotential for the meson. The red lines represent the dual marginal deformations on both sides of the duality after the addition of the respective deformations.}
\label{SD}
\end{center}
\end{figure}

Here we first revisit some well-known facts about Seiberg duality for $SU(N_c)$ SQCD 
with $N_f=2N_c$.
The electric and the magnetic theory have the same gauge group,
the electric theory has $W_\text{ele} = 0$ while in the dual theory there is a non-vanishing superpotential $W_\text{mag} = h \text{Tr} (M q \tilde q)$, where 
$M = Q \tilde Q$ is the meson of the electric theory that corresponds to an elementary singlet in the magnetic description.
This theory can be deformed on the electric side by quartic operators in the gauge  invariant combinations of $Q$ and
$\tilde Q$. These deformations are exactly marginal, they do not imply any RG flow and they just explore the conformal manifold. For example if we perturb the electric superpotential by $\Delta W_\text{ele} = \xi \text{Tr} (Q \tilde Q)^2$ such an exactly marginal deformation corresponds to one direction in the conformal manifold, emanating from the point $(g,\xi) = (g^*,0)$ (see the LHS of figure \ref{SD}).
This deformation modifies
the dual superpotential  as $\Delta W_\text{mag} = \hat \xi \text{Tr} M^2$, which
 is  marginal as well, because of Seiberg duality. Indeed for $N_f=2N_c$ the singlet $M$ has R-charge
$R_M=1$. Also in this case turning on the deformation correspond to moving on a line of conformal fixed points, emanating in this case from the point $(\tilde g, h, \tilde \xi) = (\tilde g^*, h^*, 0)$  (see the RHS of figure \ref{SD}).
In general one can extend the notion of  duality along the whole line, indeed along this line of the conformal manifold the 
two models share the same central charges, global anomalies and superconformal index.
We will refer to these types of duals as \textit{conformal duals}, borrowing the terminology of \cite{Razamat:2019vfd,Razamat:2020gcc,Razamat:2020pra}.
In this sense $SU(N_c)$ SQCD with $N_f=2N_c$ and $W =  \xi \Tr (Q \tilde Q)^2$ is \textit{conformally dual} to 
$SU(N_c)$ SQCD with $N_f=2N_c$ and $W =  h \Tr ( M q \tilde q)$.

Another crucial point of having a singlet $M$ with R-charge equal to $1$ is that this field allows us to write down a (marginal)  \emph{mass} deformation in the superpotential. By \emph{integrating out} such a mass term the theory becomes self-dual and Seiberg duality 
in this case is equivalent to S-duality.
Indeed as discussed above S-duality for $\mathcal{N}=1$ SQCD is  referred to as inherited duality \cite{Argyres:1996eh,Argyres:1999xu}, obtained by  deforming  
$\mathcal{N}=2$ SQCD with $N_f=2N_c$ by a mass term for the adjoint field in the $\mathcal{N}=2$ vector multiplet.
The deep connection between $\mathcal{N}=1$ S-duality and the presence of singlets with R-charge equal to $1$ was first observed in \cite{Leigh:1995ep} and will play a crucial role in our discussion.
%
%
%
\section{\texorpdfstring{$\mathcal{N}=2$ elliptic models}{N=2 elliptic models}}
\label{sec:elliptic}
\begin{figure}[h]
    \centering
    \includegraphics[width=0.5\textwidth]{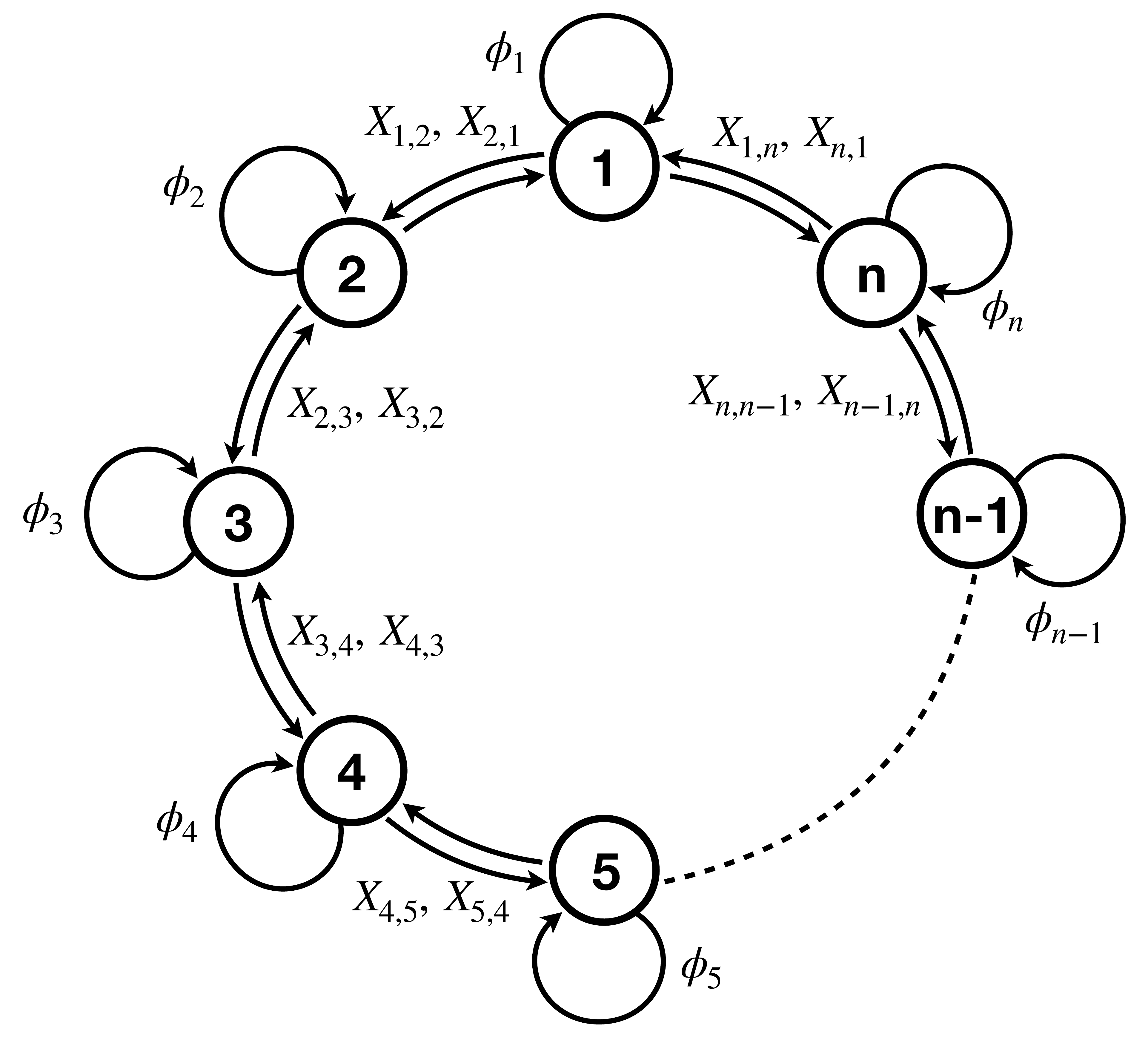}
    \caption{Quiver diagram of an $\mathcal{N}=2$ necklace theory with $n$ $SU(\nc)$ gauge groups. The numbers in the circles are used to  progressively label the $SU(\nc)$ groups starting with a `1'. We will use this convention throughout the paper.}
    \label{fig:necklace}
 \end{figure}

%
%
%
The four-dimensional $\mathcal{N}=2$ gauge theories from which we will start our analysis are known as elliptic models \cite{Witten:1997sc}. They are given by necklaces of $n$ $SU(\nc)$ gauge groups connected by bifundamental hypermultiplets
and  they can be conveniently represented using quiver diagrams. In $\mathcal{N}=1$ language, these quivers consist of $n$ nodes placed on a circle, corresponding to the $n$ gauge groups (figure \ref{fig:necklace}). Each pair of consecutive nodes is connected by two opposite arrows, $X_{i,i+1}$ and $X_{i+1,i}$ (with $i=1,\dots, n$ and with the label $n+1$ identified with the label $1$) representing two chiral multiplets in the bifundamental representation of $SU(\nc)$, which correspond to the $\mathcal{N}=2$ hypermultiplets. Furthermore,  a closed arrow representing an adjoint field $\phi_i$ is associated 
to each node.  This corresponds to the $\mathcal{N}=1$ adjoint chiral multiplet in the
 $\mathcal{N}=2$ vector multiplet.
Our conventions for representing tensorial matter in quiver diagrams are shown in table \ref{tenrosial_conventions}.  When there are two possible representations for a tensorial field (e.g. the antisymmetric representation for an orthogonal gauge group) we choose to represent it as an adjoint.

%
%
\begin{table}

	\renewcommand{\arraystretch}{2.6}
	\centering
\addtolength{\leftskip} {-2cm} 
\addtolength{\rightskip}{-2cm}
	\begin{adjustbox}{width=\columnwidth,center}
	\begin{tabular}{|l|l|}
		\hline
		\multicolumn{2}{|c|}
		{\tikz[scale=1.8,baseline=-0.5ex, thick]{ \draw[dashed] (-0.2,0) -- (-0.6,0); 
				\clip (-0.6,-0.3) rectangle (0.8,0.3);
				\draw (0,0) circle [radius=0.2]; 
				\centerarc[<-](0.3,0)(-135:135:0.2); 
				\node[anchor=west] at (0.5,0) {$\phi$} } = \text{ adjoint}}
		\\
		\hline
		\tikz[scale=1.8,baseline=-0.5ex, thick]{ \draw[dashed] (-0.2,0) -- (-0.6,0); 
			\clip (-0.6,-0.3) rectangle (0.8,0.3);
			\draw (0,0) circle [radius=0.2]; 
			\centerarc[orange](0.3,0)(-135:135:0.2); 
			\node[anchor=west] at (0.5,0) {$S$} } = \text{ symmetric }
		&
		\multirow{2}{*}{
			\tikz[scale=1.8,baseline=-0.5ex, thick]{ \draw[dashed] (-0.2,0) -- (-0.6,0); 
				\draw (0,0) circle [radius=0.2]; 
				\centerarc[orange,dashed](0.3,0)(-135:135:0.18); 
				\centerarc[orange](0.3,0)(-135:135:0.22); 
				\node[anchor=west]  at (0.5,0) {$S$, $\tilde{S}$}; } = 
			$\begin{gathered}\text{symmetric and} \\ \text{symmetric conjugate}\end{gathered}$
		}
		\\
		\tikz[scale=1.8,baseline=-0.5ex, thick]{ \draw[dashed] (-0.2,0) -- (-0.6,0); 
			\clip (-0.6,-0.3) rectangle (0.8,0.3);
			\draw (0,0) circle [radius=0.2]; 
			\centerarc[orange, dashed](0.3,0)(-135:135:0.2); 
			\node[anchor=west] at (0.5,0) {$\tilde{S}$} } = \text{ symmetric conjugate }
		&
		\\
		\hline
		\tikz[scale=1.8,baseline=-0.5ex, thick]{ \draw[dashed] (-0.2,0) -- (-0.6,0); 
			\clip (-0.6,-0.3) rectangle (0.8,0.3);
			\draw (0,0) circle [radius=0.2]; 
			\centerarc[azzurro](0.3,0)(-135:135:0.2); 
			\node[anchor=west] at (0.5,0) {$A$} } = \text{ antisymmetric }
		&
		\multirow{2}{*}{
			\tikz[scale=1.8,baseline=-0.5ex, thick]{ \draw[dashed] (-0.2,0) -- (-0.6,0); 
				\draw (0,0) circle [radius=0.2]; 
				\centerarc[azzurro,dashed](0.3,0)(-135:135:0.18); 
				\centerarc[azzurro](0.3,0)(-135:135:0.22); 
				\node[anchor=west]  at (0.5,0) {$A$, $\tilde{A}$}; } = 
			$\begin{gathered}\text{antisymmetric and} \\ \text{antisymmetric conjugate}\end{gathered}$
		}
		\\
		\tikz[scale=1.8,baseline=-0.5ex, thick]{ \draw[dashed] (-0.2,0) -- (-0.6,0); 
			\clip (-0.6,-0.3) rectangle (0.8,0.3);
			\draw (0,0) circle [radius=0.2]; 
			\centerarc[azzurro, dashed](0.3,0)(-135:135:0.2); 
			\node[anchor=west] at (0.5,0) {$\tilde{A}$} } = \text{ antisymmetric conjugate }
		&
		\\
		\hline
	\end{tabular}
	\end{adjustbox}
	\renewcommand{\arraystretch}{1}
	\caption{Conventions for representing tensorial matter in quiver diagrams. When there are two possible representations for a tensorial field (e.g. the antisymmetric representation for an orthogonal gauge group) we choose to represent it as an adjoint.}
	\label{tenrosial_conventions}
\end{table}

\subsection{S-duality and inherited dualities}
As  discussed in \cite{Witten:1997sc}  the duality group that acts on the complexified gauge couplings of these gauge quivers corresponds to
 the mapping class group of a torus with $n$ punctures, once the Type IIA brane picture is uplifted to M-theory.
 
\begin{figure}[h]
	\centering
	\includegraphics[width=0.6\textwidth]{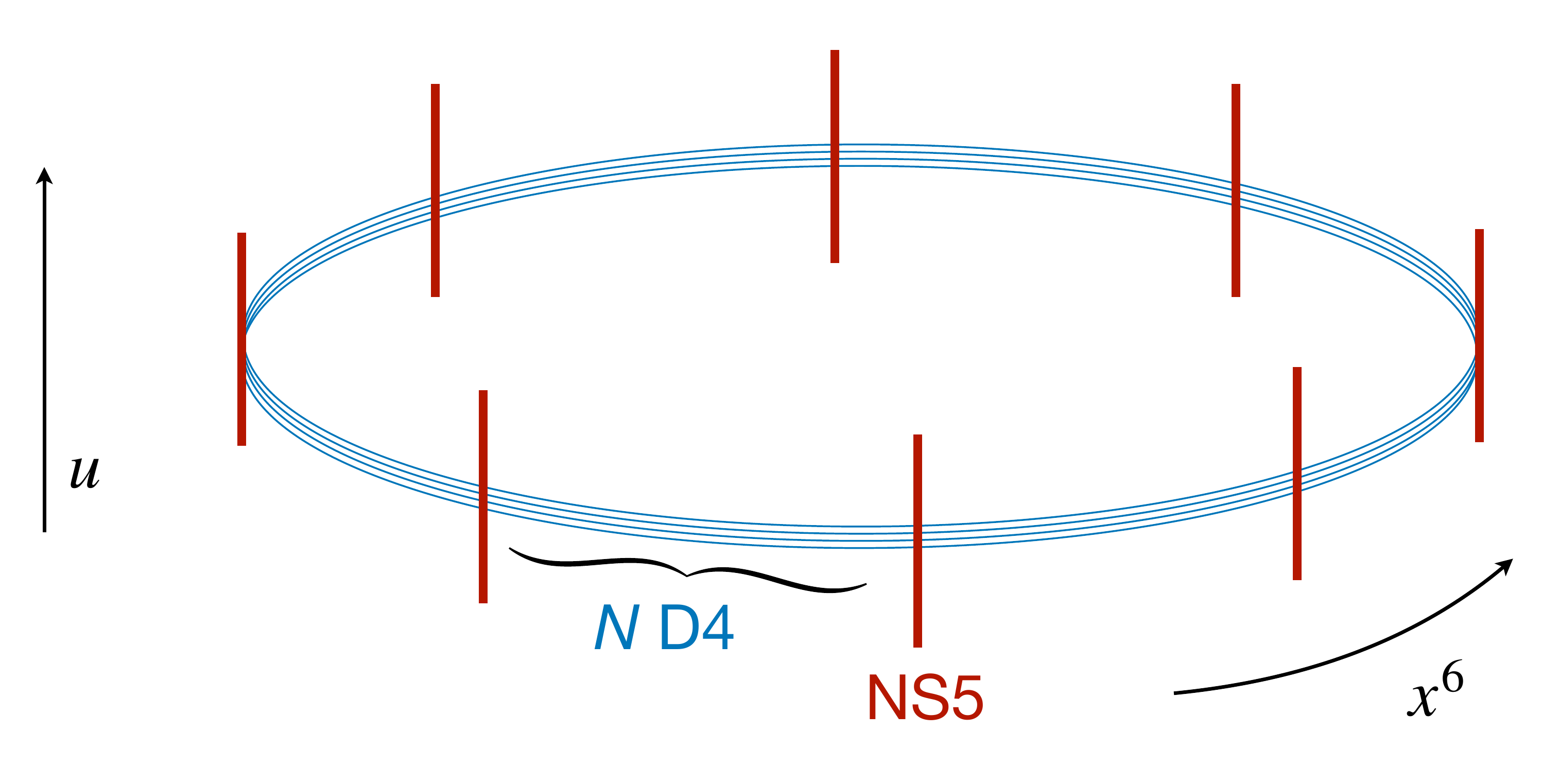}
	\caption{IIA picture of a 4d $\mathcal{N}=2$ necklace quiver with $n$ $SU(N_c)$ gauge groups: $n$ stacks of $N_c$ D4-branes suspended between $n$ distinct NS5-branes.}
	\label{fig:D4NS5}
\end{figure}

\paragraph{The $\mathcal{N}=2$ duality group.}
Here we review the duality group of $\mathcal{N}=2$  necklace quivers \cite{Witten:1997sc}.
These theories have a Type IIA description in terms of $n$ separated NS5-branes parallel to each other, with $\nc$ D4-branes stretched between each pair of nearby NS5 (figure \ref{fig:D4NS5}). Their configuration is summarized in the following table:
\[
\begin{array}{lcccccccccc}
	\toprule
	\ & 0 & 1 & 2 & 3 & 4 & 5 & 6 & 7 & 8 & 9 \\
	\midrule
	\text{D4} & - & - & - & - & \cdot & \cdot & - & \cdot & \cdot & \cdot \\
	\text{NS5} & - & - & - & - & - & - & \cdot & \cdot & \cdot & \cdot \\
	\bottomrule
\end{array}
\]
The $x^6$ direction is periodic with period $2\pi L$ and the $i$-th NS5-brane is located at $x^6 = x_i^6$. %
The IIA description can be lifted to M-theory along the compact direction $x^{10}$, which has period $\theta R$. The NS5-branes lift to M5-branes located at $n$ fixed position in $x^6$ and $x^{10}$, while the $\nc$ D4-branes lift to a single M5-branes wrapped $\nc$ times on the two compact directions $x^6$ and $x^{10}$. In other words, this construction can be thought of as a single M5-brane that wraps a complex curve embedded in two complex dimensions with coordinates $(u,w)$,  $u=x^4+i x^5$,  $w = x^6 + i x^{10}$. The global geometric picture consists of an $N_c$-cover of a two-torus with $n$ punctures. The distance between two consecutive punctures along the compact directions of the torus can be interpreted as the gauge coupling of a node of the quiver, and it is given by:
\begin{equation}
	\begin{split}
		& \tau_j = \frac{i(x^6_{j+1} - x^6_j)}{16\pi^2 g_s L} + \frac{x^{10}_{j+1} - x_j^{10}}{2\pi R}\ , \qquad  j=1,\dots,n-1\ ;\\
		& \tau_n = \frac{i(x^6_1 - x^6_n + 2\pi L)}{16 \pi^2 g_s L} + \frac{x_1^{10} - x_n^{10} + \theta R}{2\pi R}\ .
	\end{split}
	\label{gauge_couplings}
\end{equation}
The duality group acting on the gauge couplings of this class of four-dimensional quiver gauge theories is thus the mapping class group of a two-torus with $n$ punctures, also denoted as $M(1,n)$ \cite{Birman}. This group contains as subgroups $SL(2, \mathbb{Z})$, the $\hat{A}_{n-1}$ affine Weyl group and a $\mathbb{Z}_n$ that rotates the nodes of the quiver. In fact, the generators of $M(1,n)$ are the generators $S$ and $T$ of $SL(2,\mathbb{Z})$, the permutations $s_i$ and the shifts $T_i$ and $t_i$. In the gauge coupling basis $(\tau_1, \dots, \tau_n)$, they act as
\begin{equation} 	
	\begin{split}
		& S: \ (\tau_1, \dots, \tau_n) \to \biggl( \frac{\tau_1}{\tau}, \dots, \frac{\tau_{n-1}}{\tau}, \frac{\tau_n}{\tau}-1-\frac{1}{\tau} \biggr)\ ,\\
		& T: \ (\tau_1, \dots, \tau_n) \to (\tau_1, \dots, \tau_{n-1}, \tau_n+1)\ ,\\
		& T_i: \ (\tau_1, \dots, \tau_n) \to (\tau_1, \dots, \tau_{i-1}+1, \tau_i-1, \dots, \tau_n)\ ,\\    
		& t_i: \ (\tau_1, \dots, \tau_n) \to (\tau_1, \dots, \tau_{i-1}+\tau, \tau_i-\tau, \dots, \tau_n)\ ,\\      
		& s_i: \ (\tau_1, \dots, \tau_{i-1}, \tau_i, \tau_{i+1}, \dots, \tau_n) \to (\tau_1, \dots, \tau_{i-1}+\tau_i, -\tau_i, \tau_{i+1}+\tau_i, \dots, \tau_n)\ ,
	\end{split}
	\label{transf:couplings}
\end{equation}
where $\tau$ is the overall gauge coupling constant, $\tau= \sum_i \tau_i$.
\paragraph{The $\mathcal{N}=1$ inherited duality.}
The $\mathcal{N}=2$ supersymmetry of the gauge theory can be broken to $\mathcal{N}=1$ by giving masses to the adjoint scalars. In the IIA picture, this can be achieved by rotating a subset of all NS5-branes with respect to the others. (See figure \ref{fig:IIAN1} and table \ref{tab:IIAN1}.) In M-theory, this corresponds to the M5-brane wrapping a curve embedded in three complex dimensions with coordinates $(u,v,w)$, where the extra dimension is given by $v=x^7+i x^8$. As shown in \cite{Halmagyi:2004ju}, when supersymmetry is broken to $\mathcal{N}=1$ by small or large masses for the adjoints, the group $M(1,n)$ acts on these masses in a way analogous to \eqref{transf:couplings}. Therefore the $\mathcal{N}=1$ duality group is the same as the $\mathcal{N}=2$ one. 
 This is another realization of the inherited S-duality from $\mathcal{N}=2$ to $\mathcal{N}=1$ theories.
\begin{table}[h]
	\begin{minipage}{0.55\textwidth}
		\centering
		\includegraphics[width=\textwidth]{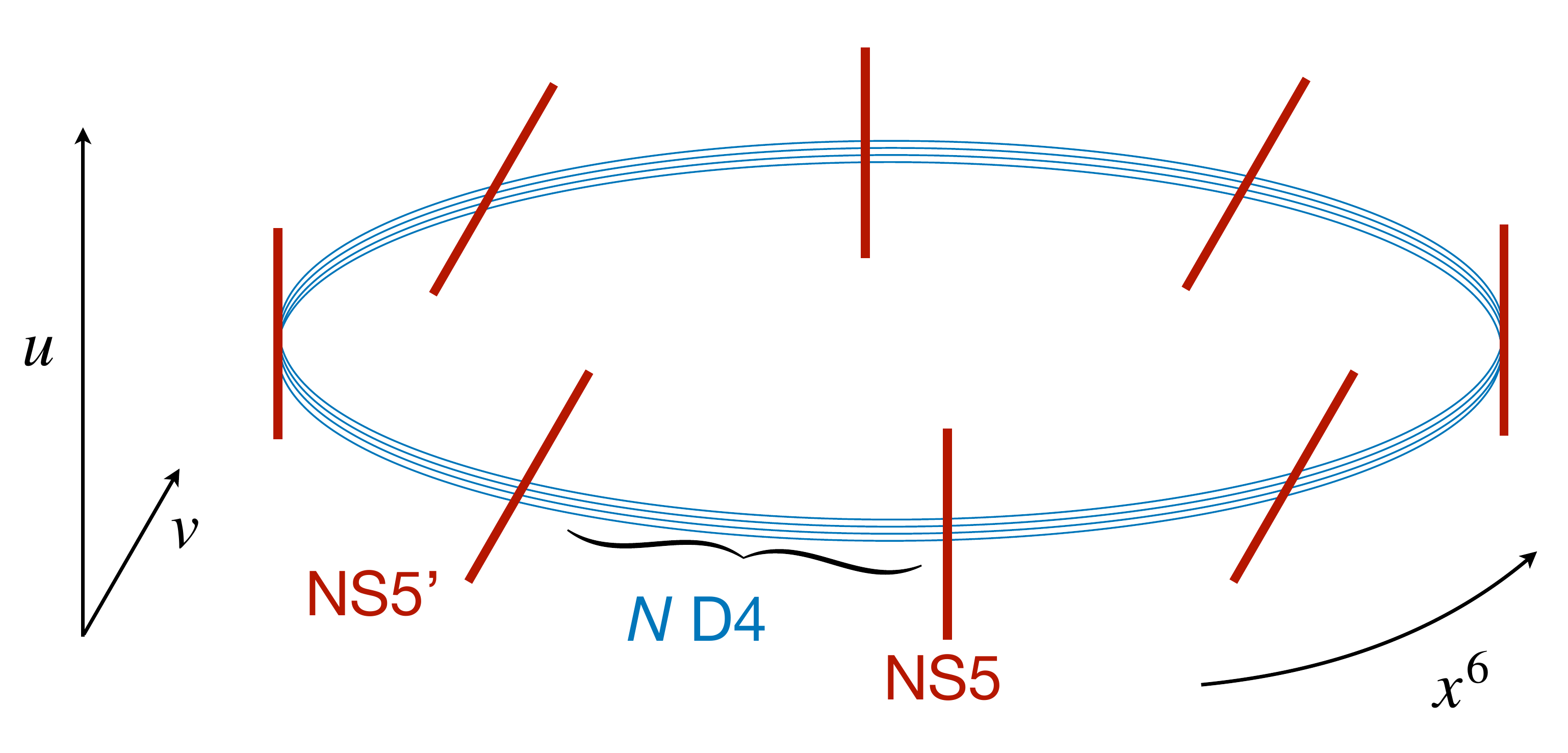}
		\captionof{figure}{IIA picture of a 4d $\mathcal{N}=1$ necklace quiver with $n$ $SU(\nc)$ gauge groups: the relative rotation between two consecutive NS5-NS5' branes breaks supersymmetry from $\mathcal{N}=2$ to $\mathcal{N}=1$.}
		\label{fig:IIAN1}
	\end{minipage}
\qquad
	\begin{minipage}{0.35\textwidth}
		\centering
		\[
		\begin{array}{lcccccccccc}
			\toprule
			\ & 0 & 1 & 2 & 3 & 4 & 5 & 6 & 7 & 8 & 9 \\
			\midrule
			\text{D4} & - & - & - & - & \cdot & \cdot & - & \cdot & \cdot & \cdot \\
			\text{NS5} & - & - & - & - & - & - & \cdot & \cdot & \cdot & \cdot \\
			\text{NS5'} & - & - & - & - & \cdot & \cdot & \cdot & - & - & \cdot \\
			\bottomrule
		\end{array}
		\]
		\caption{IIA configuration configuration of D4-branes, NS5-branes and NS5'-branes.}
		\label{tab:IIAN1}
	\end{minipage}
\end{table}

\subsection{\texorpdfstring{$\mathcal{N}=2$ S-dualities from orientifold projections}{N=2 S-dualities from orientifolds projections}}
Quivers with orthogonal and symplectic gauge groups can be obtained from the elliptic models described in the previous section by introducing orientifold sixplanes (O6). The inclusion of 
 O6$^-$-planes in the $(0123789)$ directions does not break further supersymmetry, so the resulting low-energy effective theories on the D4 worldvolume will have $\mathcal{N} = 2$ supersymmetry. These configurations were realized in \cite{Uranga:1998uj}. Here we consider pairs of orientifolds placed at antipodal points in the compact $x^6$ direction, say at $x^6=0$ and $x^6=L/2$. Moreover we restrict to pairs of orientifolds with opposite RR charges; configurations with orientifolds with the same charge are possible, but they require additional D6-branes (flavors) in order for the $\beta$ function to vanish. In this paper we do not consider theories with flavor.

When an O6  is placed on a stack of $2 N_c$ D4-branes the gauge group is projected to $SO(2N_c)$ for positive orientifold charge (O6$^+$) or to $USp(2N_c)$ for negative orientifold charge (O6$^-$). When an  O6  is placed on top of an NS5-brane that separates two stacks of $N_c$ D4-branes the two $SU(N_c)$ gauge groups are identified and we find matter in a tensorial representation (in the symmetric of $SU(N_c)$ for  O6$^+$  and in the antisymmetric for  O6$^-$). The brane configuration needs to be symmetric with respect to the reflection induced by the orientifold planes.

We consider the configurations with one O6 placed on a stack of D4's and the other placed on an NS5 with opposite RR charge; these correspond to the families \textbf{ii)} and \textbf{iii)} of \cite{Uranga:1998uj}. The configuration with $2n_g+1$ NS5's and the O6$^-$ on top of an NS5 has a low-energy effective worldvolume theory given by the following $\mathcal{N}=2$ quiver:\footnote{We will always use $\nn=1$ notation when representing quiver diagrams, even when we discuss theories with extended supersymmetry.}
\begin{equation}	\label{eq:N=2_SO_quiver}
	\QuiverNtwo{$SO(2N_c+1)$}{$SU(2N_c-1)$}{$SU(2N_c-3)$}{$SU(2(N_c-n_g) + 1)$}{$A,\tilde{A}$}{0.8}{}
\end{equation}
with bifundamental hypermultiplets between nearby groups and an ($\mathcal{N}=2$) antisymmetric hypermultiplet for the rightmost $SU$ group. This is model \textbf{ii)} considered in \cite{Uranga:1998uj} with odd ranks for the gauge groups.  We will refer to it as the $\mathcal{N}=2$ model $\mathcal{A}$. The ranks have been chosen in order to set all the $\beta$ functions to zero.  (Moreover we must have $N_c>n_g$.) In \cite{Uranga:1998uj} it was shown that this Type IIA configuration can be uplifted  to a single smooth M5-brane in M-theory. This M5-brane is a $(2N_c+1)$-fold cover of the M-theory torus parameterized by $x^6$ and $x^{10}$ with $2n_g+1$ punctures and is consistent with the $\zz_2$ quotient induced by the orientifolds. It was also shown that only a $2N_c$-fold cover submanifold of the M5-brane is dynamical.

Next we consider the configuration with $2n_g+1$ NS5's and the O6$^+$ on top of an NS5. The corresponding $\nn=2$ quiver is:
\begin{equation}	\label{eq:N=2_USp_quiver}
	\QuiverNtwo{$SU(2N_c)$}{$SU(2N_c-2)$}{$SU(2N_c-4)$}	{$USp(2(N_c-n_g))$}{}{0.8}{$S,\tilde{S}$}
\end{equation}
with bifundamental hypermultiplets between nearby groups and an ($\mathcal{N}=2$) symmetric hypermultiplet for the leftmost $SU$ group. Again, all the $\beta$ functions vanish with this rank assignment (and $N_c>n_g$).  We will refer to this model as the $\mathcal{N}=2$ model $\mathcal{B}$.
In \cite{Uranga:1998uj} it was shown that this configuration (model \textbf{iii)} there) can be uplifted to M-theory. In M-theory this model is described by a single smooth M5-brane. The M5-brane is a $2N_c$-fold cover of the M-theory torus with $2n_g+1$ punctures consistent with the $\zz_2$ quotient induced by the orientifolds.

In \cite{Uranga:1998uj} it was claimed that these two models are S-dual. Indeed it was shown that the two Type IIA elliptic models can be obtained as different classical limits of the same M-theory configuration, namely by shrinking different cycles of the M-theory torus.

A straightforward  extension of the notion of S-duality for these models is given by the  $SL(2,\zz)$ modular group of the M-theory torus. 
The full S-duality group is related to the isomorphism group of the torus with $2n_g+1$ punctures modulo the $\zz_2$ quotient induced by the orientifolds, though further analysis is needed to determine the full S-duality action.

Observe that these quiver gauge theories can be equivalently engineered in a brane tiling picture.
The general analysis has been recently performed  in \cite{Antinucci:2021edv}, 
by following the recipe of \cite{Franco:2007ii} (see also \cite{Argurio:2020dko} for recent updates on the prescription).
All of the models considered here can be studied  with this technique, by including fractional branes and orientifolds in the brane tiling.  Nevertheless we will not make use of this construction here.

%
%
%
%
%
\section{\texorpdfstring{New $\mathcal{N}=1$ inherited S-dualities}{New N=1 inherited  S-dualities}}
\label{sec:newind}
%
%
%
%
\subsection{A general classification}
In this section we  generalize the classification of the elliptic models considered above by  considering rotated NS5-branes (NS5') and O6-planes (O6'). Some of these models (corresponding to pairs of O6 and pairs of O6'-planes) were studied in \cite{Uranga:1998uj,Park:1999eb} where a Type IIB T-dual description was also constructed. The directions wrapped by the branes are summarized in the following table:
\[
\begin{array}{lcccccccccc}
	\toprule
	\ & 0 & 1 & 2 & 3 & 4 & 5 & 6 & 7 & 8 & 9 \\
	\midrule
	\text{D4} & - & - & - & - & \cdot & \cdot & - & \cdot & \cdot & \cdot \\
	\text{NS5} & - & - & - & - & - & - & \cdot & \cdot & \cdot & \cdot \\
	\text{NS5'} & - & - & - & - & \cdot & \cdot & \cdot & - & - & \cdot \\
	\text{O6} & - & - & - & - & \cdot & \cdot & \cdot & - & - & - \\	
	\text{O6'} & - & - & - & - & - & - & \cdot & \cdot & \cdot & - \\
	\bottomrule
\end{array}
\]
\begin{figure}[t]
	\centering
	\includegraphics[width=0.7\textwidth]{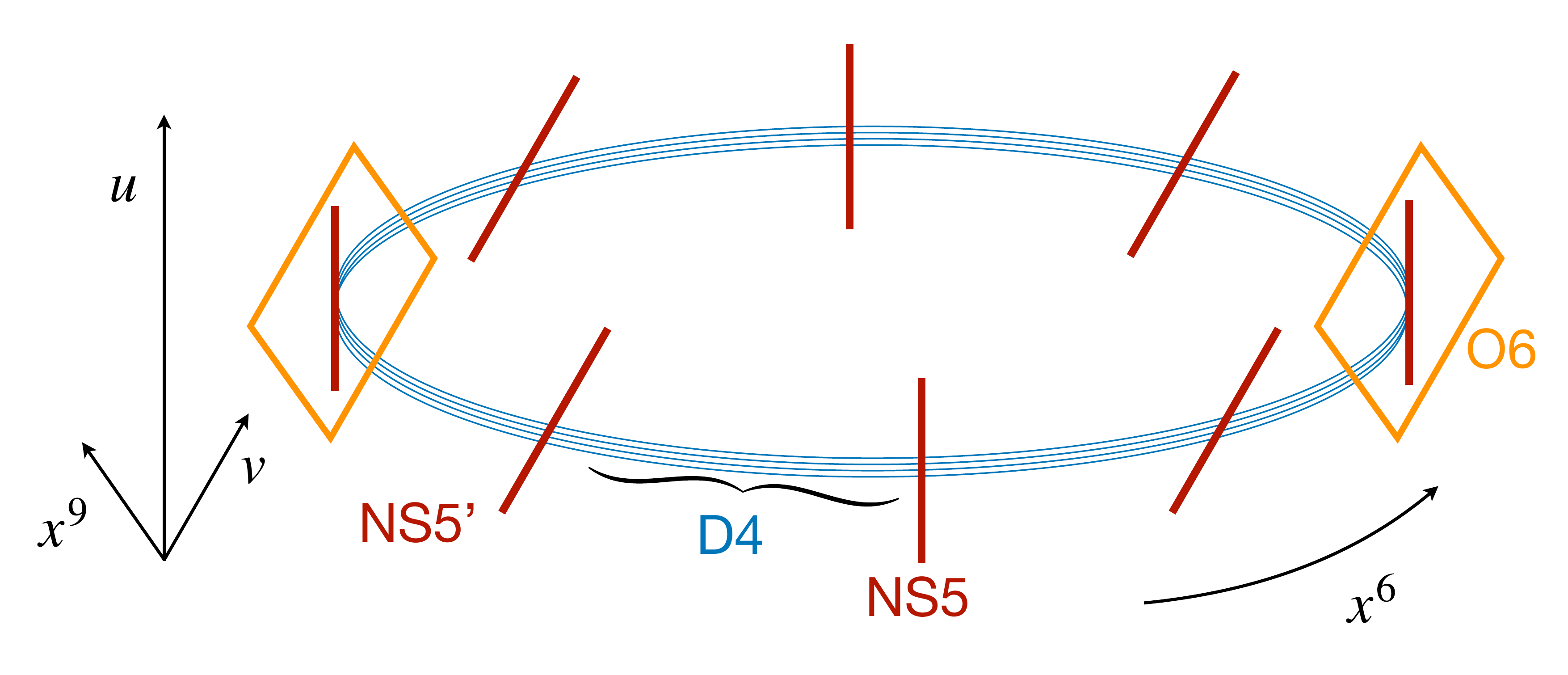}
	\caption{IIA picture of an elliptic model with O6-planes. The low-energy effective quiver theory can have orthogonal and/or symplectic gauge groups.}
	\label{fig:D4NS5O6}
\end{figure}%
The presence of both primed and non-primed branes breaks supersymmetry to $\mathcal{N}=1$: a pair of nearby NS-NS' gives infinite mass to the scalar in the $\mathcal{N}=2$ vector multiplet breaking supersymmetry to $\mathcal{N}=1$, while an NS-O6'-NS (NS'-O6-NS') configuration brings $\mathcal{N}=1$ chiral multiplets in conjugate tensorial representations other than the adjoint. Again we do not consider models which need additional flavor D6-branes. NS5-branes must be tilted in pairs, according to the $\zz_2$ quotient. A possible configuration is shown in figure \ref{fig:D4NS5O6}.

The  classification of possible configuration of O6-planes and NS5-branes (coincident with the O6-planes) is shown in table \ref{tab:classification}. There, the resulting quiver theories are shown schematically with the leftmost and rightmost gauge groups together with their tensorial matter. The other nodes have unitary gauge groups with (without) adjoint if the corresponding NS5-branes are parallel (non-parallel). Additionally, the real groups may have either an adjoint, or a tensorial representation other than the adjoint or may not have any tensorial matter at all. The type of tensorial matter for the real groups depends on the orientation of the corresponding O6-plane, while the presence of tensorial matter depends on the orientation of the two nearest NS5-branes. Quivers without tensorial matter for the real groups can be engineered by considering O6-planes at an angle in the (4578) space, but we do not consider this possibility here.

Following the notation of \cite{Uranga:1998uj}, there are four families of quivers distinguished by the leftmost and rightmost gauge group. In the following we consider configurations which correspond to quivers with $n_g+1$ nodes. The amount of NS5-branes needed to obtain such theories depends on the family.\footnote{For family \emph{i)} one needs $2n_g$ NS5's, for families \emph{ii)} and \emph{iii)} one needs $2n_g+1$ NS5's, and for family \emph{iv)} one needs $2n_g+2$ NS5's. Furthermore observe that the number of regular D4-branes corresponds to $2(N_c-n_g)$ for family  \emph{i)} and \emph{ii)},  $2(N_c-n_g)+1$  for family \emph{iii)}, and  $N_c-2n_g$ for family \emph{iv)}.} The $i$-th node has an adjoint if $i\in I$,  where $I$ is a set of indices (which is most easily defined in explicit examples -- see section \ref{sec:exam}), while nodes with $i\notin I$ do not have an adjoint. For the sake of simplicity we have not drawn the adjoints. The enumeration of the nodes in the quivers goes from left to right, starting at 1.

\begin{itemize}[leftmargin=0.5cm]
\item[\emph{i)}] This is realized  without NS5-branes intersecting the O-planes. The quiver is:
\begin{equation}
	\Quiver{$SO(2N_c)$}{$SU(2N_c-2)$}{$SU(2N_c-4)$}{$USp(2(N_c-n_g))$}{}{0.8}{}
\end{equation}
with superpotential
\begin{equation}
	\begin{split}
		W =&\ \phi_1 X_{12} X_{21} + \sum_{i\in I} \phi_i (X_{i,i-1}X_{i-1,i} - X_{i,i+1}X_{i+1,i}) 
		\\&+ \sum_{i\notin I} (\pm X_{i,i-1}X_{i-1,i}X_{i,i+1}X_{i+1,i}) + \phi_{n_g+1} X_{n_g+1,n_g} X_{n_g,n_g+1}\ ,
	\end{split}
\end{equation}
where  traces are understood as well as the correct sums over the color indices in each term. The $i$-th node has an adjoint iff $i\in I$. $\mathcal{N}=2$ models in this family, corresponding to cases with parallel NS5-branes,  are self-dual under S-duality, 
i.e. the S-dual phases have the same gauge algebra and matter content.  

\item[\emph{ii)}] This is realized when an NS5 intersects the O6$^-$ and there are no NS5's intersecting the O6$^+$. The quiver is:
\begin{equation}		\label{eq:N=1_SO_quiver}
	\Quiver{$SO(2N_c+1)$}{$SU(2N_c-1)$}{$SU(2N_c-3)$}{$SU(2N_c-2n_g+1)$}{$A,\tilde{A}$}{0.8}{}
\end{equation}
with two antisymmetric $A$, $\tilde{A}$ for the rightmost gauge group and superpotential
\begin{equation}
	\begin{split}
		W =&\ \phi_1 X_{12} X_{21} + \sum_{i\in I} \phi_i (X_{i,i-1}X_{i-1,i} - X_{i,i+1}X_{i+1,i}) 
		\\&+ \sum_{i\notin I} (\pm X_{i,i-1}X_{i-1,i}X_{i,i+1}X_{i+1,i}) + \phi_{n_g+1} A \tilde{A}\ .
	\end{split}
\end{equation}
where the $i$-th node has an adjoint iff $i\in I$.  We will refer to this family as theory $\mathcal{A}$.

\item[\emph{iii)}] This is realized when an NS5 intersects the O6$^+$ and there are no NS5's intersecting the O6$^-$. The quiver is:
\begin{equation}	\label{eq:N=1_USp_quiver}
	\Quiver{$SU(2N_c)$}{$SU(2N_c-2)$}{$SU(2N_c-4)$}	{$USp(2(N_c-n_g))$}{}{0.8}{$S,\tilde{S}$}
\end{equation}
with two conjugate symmetric tensors  $S$ and $\tilde{S}$ for the leftmost gauge group and superpotential
\begin{equation}
	\begin{split}
		W =&\ \phi_1 S \tilde{S} + \sum_{i\in I} \phi_i (X_{i,i-1}X_{i-1,i} - X_{i,i+1}X_{i+1,i}) 
		\\&+ \sum_{i\notin I} (\pm X_{i,i-1}X_{i-1,i}X_{i,i+1}X_{i+1,i}) + \phi_{n_g+1} X_{n_g,n_g+1} X_{n_g+1,n_g}\ .
	\end{split}
\end{equation}
where the $i$-th node has an adjoint iff $i\in I$. We will refer to this family as theory $\mathcal{B}$.  S-duality maps model $\mathcal{A}$ to model  $\mathcal{B}$ in the $\mathcal{N}=2$ case.  

\item[\emph{iv)}]  This is realized when both O6$^-$-planes have an NS5 on top. The quiver is:
\begin{equation}
	\Quiver{$SU(N_c)$}{$SU(N_c-2)$}{$SU(N_c-4)$}{$SU(N_c-2n_g)$}{$A,\tilde{A}$}{0.8}{$S,\tilde{S}$}
\end{equation}
with two antisymmetric $A$, $\tilde{A}$ for the rightmost gauge group and two symmetric $S$, $\tilde{S}$ for the leftmost gauge group. The superpotential is
\begin{equation}
	\begin{split}
		W =&\ S \tilde{S} \phi_1 + \sum_{i\in I} \phi_i (X_{i,i-1}X_{i-1,i} - X_{i,i+1}X_{i+1,i}) 
		\\&+ \sum_{i\notin I} (\pm X_{i,i-1}X_{i-1,i}X_{i,i+1}X_{i+1,i}) + \phi_{n_g+1} A \tilde{A}\ ,
	\end{split}
\end{equation}
where the $i$-th node has an adjoint iff $i\in I$. 
In this case the models are not in general self-dual under S-duality.\footnote{We are grateful to I.~Garc\'ia-Etxebarria for comments on this point.}
For example when $n_g=0$ the S-dual models have been found in 
\cite{Chacaltana:2012zy,Chacaltana:2014nya}. However for higher values of $n_g$ one can still have shift and permutation symmetries for the NS5-branes that are not placed on top of the O-planes.
\end{itemize}
(In the above survey we have described the case where the leftmost and rightmost groups have an adjoint $\phi_1$, $\phi_{n_g+1}$, or either an adjoint or a tensor for the real groups. If they do not have such two-index matter field, the corresponding superpotential term must be replaced with a quartic term, in a straightforward way.)

We claim that the notion of S-duality for the  $\mathcal{N}=1$  models in families \emph{ii)} and \emph{iii)}, i.e.  model $\mathcal{A}$ and $\mathcal{B}$ respectively where some of the NS5-branes are non-parallel, is inherited from the S-duality of the $\mathcal{N}=2$ case (where all NS5's are parallel).
Furthermore we claim that a generalized notion of inherited S-duality holds in the $\mathcal{N}=1$ case  for all models  summarized above.
This duality requires moving on the conformal manifold of the two theories, and for this reason we dub it \textit{conformal duality}.
We will exhibit this duality only for the models in families \emph{ii)} and \emph{iii)}, but this result is more general and it is related to the 
presence of adjoint fields with R-charge equal to $1$ in the $\mathcal{N}=1$ case.
In the following we will perform various explicit checks of the inherited duality between model $\mathcal{A}$ and model $\mathcal{B}$. All checks for the other models are  
instead straightforward, because they differ only by adjoints with R-charge  $1$ that do not contribute to the anomalies, to the central charges or to the superconformal index. (Indeed the coefficients of the Weyl and of the Euler density for a superfield with R-charges equal to $1$ are vanishing, and the contribution to the 
superconformal index is $1$.)
For family \emph{i)} this observation has already been made in \cite{Antinucci:2021edv}.  For family \emph{iv)} the above discussion does not apply for the cases that involve NS5-branes placed on top of the O-planes. Indeed in those cases we cannot 
break supersymmetry to $\mathcal{N}=1$ without modifying the model by the addition of extra D6-branes. This is consistent with the fact that the model for $n_g=0$ is not self-dual under S-duality.


\begin{landscape}
	\begin{table} 
		\centering
		\begin{tabular}{c|c|c|c|c|c}%
			\multicolumn{4}{c}{} & \multicolumn{1}{c}{Theory $\mathcal{B}$} &  \multicolumn{1}{c}{Theory $\mathcal{A}$}\\
			\renewcommand{\arraystretch}{3.5}
			$
			\begin{gathered}
				\tikz[scale=0.4] 	\draw[mandarancio] (-1.8,-0.5) -- (-1.8,.4) -- (-1.2,.5) -- (-1.2,-.4) -- (-1.8,-0.5);	= \text{O6}\\
				\tikz[scale=0.4]	\draw[terrabruciatadiconcorezzo, thick] (-1.5,-0.45) -- (-1.5,0.45);	= \text{NS5}\\
				\tikz[scale=0.4,baseline=-0.5ex]	\draw[azzurro, thick] (0,0) -- (1,0);	= \text{D4} \\
			\end{gathered}
			$
			& $\mathcal{N}$ & \SimpElliptic{NS}{NS} & \SimpElliptic{}{} & \SimpElliptic{NS}{} & \SimpElliptic{}{NS} \\ 
			\hline
			O6$^+$-O6$^-$ & 2 &\LeftRightQuiver{SU}{SU}{$S,\tilde{S}$}{$A,\tilde{A}$}& \LeftRightQuiver{SO}{USp}{$\phi$}{$\phi$} & \LeftRightQuiver{SU}{USp}{$S,\tilde{S}$}{$\phi$}\tikzmark{a} & \tikzmark{b}\LeftRightQuiver{SO}{SU}{$\phi$}{$A,\tilde{A}$} \\
			\hline
			O6$^+$-O6'$^-$ & 1  & flavor & \LeftRightQuiver{SO}{USp}{$\phi$}{$A$} & \LeftRightQuiver{SU}{USp}{$S,\tilde{S}$}{$A$}\tikzmark{c} & flavor \\
			\hline
			O6'$^+$-O6$^-$ & 1 & flavor & \LeftRightQuiver{SO}{USp}{$S$}{$\phi$} & flavor & \tikzmark{d}\LeftRightQuiver{SO}{SU}{$S$}{$A,\tilde{A}$} \\
			\hline
			O6'$^+$-O6'$^-$ & 1 & flavor & \LeftRightQuiver{SO}{USp}{$S$}{$A$} & flavor & flavor \\
		\end{tabular}
		\renewcommand{\arraystretch}{1}
		\caption{Elliptic models with O6-planes and corresponding quivers.
			Models connected by \protect\tikz[baseline=-0.5ex] \protect\draw [<->, ultra thick, red] (0,0) -- (0.7,0); are S-dual.
			Configurations denoted by `flavor' need additional flavor hypermultiplets in order to make all the $\beta$ function vanish.
			The dashed line in the middle of the quiver corresponds to the intermediate $n_g-1$ nodes with unitary gauge groups.
			In this table we have assumed that all the NS5-branes are in the $(012345)$ directions; the generalization to non-parallel branes is straightforward.
			Models in the first row were considered 
			in \cite{Uranga:1998uj}; models in the fourth row were considered in \cite{Park:1999eb}.}
		\label{tab:classification}
		\begin{tikzpicture}[overlay, remember picture, shorten >=.5pt, shorten <=.5pt, transform canvas={yshift=.25\baselineskip}]
			\draw [<->, ultra thick, red] ([xshift=-5]{pic cs:a}) [bend left] -- ([xshift=4]{pic cs:b}) node[midway,above,yshift=5,fill=white] {\text{S-dual}};
			\draw [<->, ultra thick, red] ([yshift=-10]{pic cs:c})  -- ([yshift=10]{pic cs:d}) node[midway,above,yshift=5,sloped,fill=white] {\text{S-dual}};
		\end{tikzpicture}
	\end{table}
\end{landscape}

\subsection{Inherited duality for quiver theories}
We argue that S-duality between $\mathcal{N}=1$ quiver theories, namely the one relating \eqref{eq:N=1_SO_quiver} and \eqref{eq:N=1_USp_quiver}, can be seen as an inherited duality from the corresponding $\mathcal{N}=2$ models analyzed in \cite{Uranga:1998uj}. More generally, the full S-duality group of the $\mathcal{N}=2$ duality should be preserved when breaking supersymmetry to $\mathcal{N}=1$. We can break half of the supersymmetry in two  ways. 
\begin{itemize}
\item
The first  way consists in adding a mass deformation for one of the adjoints in the superpotential. This deformation is relevant and causes an RG flow to a new superconformal fixed point with $\mathcal{N}=1$ supersymmetry. In the IIA (or M-theory) picture this can be described as tilting NS5-branes. All of the quivers considered in this paper with an adjoint for the real group can be obtained in this way. 
\item
The second way consists in tilting an O6-plane.
The resulting models have a tensorial representation $T$ for the real group other than the adjoint (i.e. symmetric in the symplectic case and antisymmetric in the orthogonal one). 
This quiver theory sits on the same conformal manifold as the theory without tensorial matter for the real group because $T$ has R-charge $1$. On the other hand the theory without tensorial matter for the gauge group is obtained from the $\nn=2$ theory by giving a large mass to the adjoint of the real group. Therefore the model with tensorial matter other than the adjoint for the real group can be obtained from the $\nn=2$ theory by an RG flow followed by a motion on the conformal manifold. Schematically:
\begin{equation*}
	\tikz[scale=1.8,baseline=0ex, thick]{ \draw[dashed] (-0.2,0) -- (-0.6,0); 
		\clip (-0.6,-0.3) rectangle (0.8,0.6);
		\draw (0,0) circle [radius=0.2]; 
		\centerarc[<-](0.3,0)(-135:135:0.2); 
		\node[anchor=west] at (0.5,0) {$\phi$};
		\node[anchor=south] at (0,0.2) {SO/USp} }
	\underset{m\phi^2}{
	\quad\overset{\text{RG-flow}}{\tikz[baseline] \draw[->,thick] (0,0) -- (1.5,0);}\quad
	}
	\tikz[scale=1.8,baseline=0ex, thick]{ \draw[dashed] (-0.2,0) -- (-0.6,0); 
		\clip (-0.6,-0.3) rectangle (0.45,0.6);
		\draw (0,0) circle [radius=0.2]; 
				\node[anchor=south] at (0,0.2) {SO/USp}}
	\overset{\text{marginal}}{
	\quad\overset{\text{deformation}}{\underset{\tilde{m} T^2}{\tikz[baseline] \draw[<->,thick] (0,0) -- (1.7,0);}}\quad
	}
	\tikz[scale=1.8,baseline=0ex, thick]{ \draw[dashed] (-0.2,0) -- (-0.6,0); 
		\clip (-0.6,-0.3) rectangle (0.8,0.6);
		\draw (0,0) circle [radius=0.2]; 
		\centerarc[orange](0.3,0)(-135:135:0.2); 
		\node[anchor=west] at (0.5,0) {$T$};
		\node[anchor=south] at (0,0.2) {SO/USp} }
\end{equation*}
\end{itemize}
Here we discuss the properties of the superconformal fixed point for these $\mathcal{N}=1$ quiver theories.
The condition that the $\beta$ function vanishes at the superconformal fixed point for the inner $SU_i$ nodes reads:
\begin{equation}
	N_i(1+\eta_{\phi_i}(R_{\phi_i} -1) ) + N_{i-1}(R_{X_{i,i-1}} - 1) + N_{i+1}(R_{X_{i,i+1}} -1) = 0\ ,
\end{equation}
where $\eta_{\phi_i} = 1$ if the $i$-th adjoint is massless and vanishes otherwise. We have assumed that $X_{i,i-1}$ and $X_{i-1,i}$ have the same superconformal R-charge. This is justified by the fact that the non-R symmetries do not mix with the R-symmetry.\footnote{For the families studied here, this can be seen explicitly (using the argument of \cite{Bertolini:2004xf,Butti:2005vn}) by imposing the constraints from the anomalies and $\beta$ functions on the coefficient of the Euler density. The dependence on the mixing parameters for the non-R symmetries is quadratic and not cubic in the central charge, giving rise to linear equations in the former when maximizing the latter. This implies that these mixing coefficients can be absorbed in a redefinition of the charges and do not mix with the R-symmetry.}
The vanishing of the $\beta$ function for the rightmost group reads:
\begin{equation}
	N_{n_g+1}(1+\eta_{\phi_{n_g+1}}(R_{\phi_{n_g+1}} -1) ) + N_{n_g}(R_{X_{n_g+1,n_g}} - 1) + \frac{N_{n_g+1}+2}{2}(R_{A} -1) = 0
\end{equation}
if the two tensorial fields are antisymmetric (i.e. in theory $\mathcal{A}$), or
\begin{equation}
	N_{n_g+1}(1+\eta_{\phi_{n_g+1}}(R_{\phi_{n_g+1}} -1) ) + N_{n_g}(R_{X_{n_g+1,n_g}} - 1) + \frac{N_{n_g+1}-2}{2}(R_{S} -1) = 0
\end{equation}
if the two tensorial fields are symmetric (i.e. in theory $\mathcal{B}$). The vanishing of the $\beta$ function for the orthogonal group implies
\begin{equation}
	(2N_c-1) (1+\eta_{\phi_{1}}(R_{\phi_1} -1) ) + (2N_c-1)(R_{X_{1,2}} - 1) = 0\ ,
\end{equation}
while for the symplectic group
\begin{equation}
	(2(N_c-n_g)+2) (1+\eta_{\phi_{1}}(R_{\phi_1} -1) ) + (2(N_c-n_g)+2)(R_{X_{1,2}} - 1) = 0\ .
\end{equation}
There is only one solution for the R-charges if 
supersymmetry is broken to $\mathcal{N}=1$ and if we consider the gauge rank assignments \eqref{eq:N=1_SO_quiver} and \eqref{eq:N=1_USp_quiver}:\footnotemark
\begin{equation}	\label{eq_R-charge_assig}
	R_{\phi_i} = 1\ ,
\qquad
	R_{X_{i,i-1}} = \frac{1}{2}\ ,
\qquad
	R_{A} = R_{S} = \frac{1}{2}\ .
\end{equation}%
\footnotetext{As already discussed in the context of the IIA brane picture, the gauge anomalies vanish with this rank assignment.}%
In \cite{Antinucci:2021edv}, when discussing some of the models considered in this paper, emphasis was put on the fact that the superconformal R-charges are not obtained via $a$-maximization; rather, \eqref{eq_R-charge_assig} is the only charge assignment consistent with vanishing gauge anomalies and $\beta$ functions. Here we also emphasize that the presence of fields with R-charge equal to 1 implies the existence of marginal mass deformations for those fields, namely $m_i \phi_i^2$. 
 It was shown in \cite{Green:2010da} that a set of marginal deformations is exactly marginal if:
	\begin{equation}
		D^a = 2 \pi^2 \lambda_i T^a_{i\bar{j}} \overline{\lambda}_{\bar{j}} + \dots = 0
	\end{equation}
where $T^a$ are the generators of the non-R global symmetries and $\lambda_i$ are the coefficients of the marginal deformations. 
In our theories the possible marginal deformations are schematically:
\begin{equation}
	\phi^2\ ,
\qquad
	\phi X X\ ,
\qquad
	X X X X\ ,
\end{equation}
and similarly with $X$ exchanged with either $A$ or $S$.
These operators have vanishing charge under the non-R global symmetries (see table \ref{tab:baryonic}). Notice that this happens because the non-R global symmetries are baryonic-like, so the fields $X_{i,i+1}$ and $X_{i+1,i}$ always have opposite charge under them. We conclude that $D^a=0$ to leading order for all $a$, therefore all marginal deformations $m_i \phi_i^2$ are exactly marginal.
They correspond to directions on the conformal manifold, therefore two quivers differing by the masses of the adjoints are \textit{conformally dual}  as long as they are $\mathcal{N}=1$ supersymmetric. Similarly, if we give large masses to some of the adjoints we can integrate them out, thus two quivers differing by the number of adjoints are \textit{conformally dual} as well. The dualities considered in \cite{Antinucci:2021edv} are of this type.

The $a$ central charge of the quivers in class $\mathcal{A}$ is
\begin{equation}
	a_{\mathcal{A}} =  \frac{ 27}{64} N_c^2\left(n_g+\frac{1}{2}\right)- N_c \frac{27}{128}  \left(4 n_g^2-1\right)+\frac{9}{32} \left(n_g^3-n_g\right)
\end{equation}
and is trivially independent from the number of adjoints with R-charge 1. The central charge for the quivers in class $\mathcal{B}$ is
\begin{equation}
	a_{\mathcal{B}} =  \frac{ 27}{64}  N_c^2 \left(n_g+\frac{1}{2}\right)- N_c \frac{27}{128}  \left(4 n_g^2-1\right)+\frac{9}{32} \left(n_g^3-n_g\right)\ .
\end{equation}
The matching between $a_\mathcal{A}$ and $a_\mathcal{B}$ provides the first check of the duality between the corresponding quiver theories. Similarly it is straightforward to check that the $c$ central charges match:
\begin{equation}
	c_{\mathcal{A}} = c_{\mathcal{B}} = \frac{ 27}{64}  N_c^2 \left(n_g+\frac{1}{2}\right)- N_c \frac{27}{128}  \left(4 n_g^2-1\right)+\frac{9}{32} \left(n_g^3-n_g\right)\ .
\end{equation}
The mapping of the other global symmetries is more subtle. The only (non-R) global symmetries are abelian and baryonic-like, meaning that conjugate fields are charged oppositely and the adjoints are neutral under them. For a quiver with $n_g+1$ nodes there are $n_g+1$ such symmetries; the map of charges across the duality is given in table \ref{tab:baryonic}. We checked that the 't Hooft anomalies match using this map.
\begin{table}
	\[
\begin{array}{c|ccccccc}
	\hline
	\toprule
	\textbf{Theory $\mathcal{A}$} \ & U(1)_R & U(1)_B & U(1)_2 & U(1)_3 & U(1)_4 & \dots & U(1)_{n_g+1}\\
	\midrule
	X_{12}      & 1/2 &  1 &   2\nc-2  & 0 & 0  & \dots & 0 \\
	X_{21}      & 1/2 & -1 & -(2\nc-2) & 0 & 0 & \dots & 0 \\
	X_{23}      & 1/2 & 1  & -(2\nc-2) &   2\nc-4   & 0 & \dots & 0 \\
	X_{32}      & 1/2 & -1 &   2\nc-2  & -(2\nc-4)  & 0 & \dots & 0 \\
	X_{34}      & 1/2 &  1 & 0 & -(2\nc-4) &  2\nc-6    & \dots & 0 \\
	X_{43}      & 1/2 & -1 & 0 &   2\nc-4  & -(2\nc-6)  & \dots & 0 \\
	X_{45}      & 1/2 &  1 & 0 & 0 & -(2\nc-6) & \dots & 0 \\
	X_{54}      & 1/2 & -1 & 0 & 0 &   2\nc-6   & \dots & 0 \\
	\vdots      & \vdots & \vdots  & \vdots & \vdots & \vdots & \vdots & \vdots\\ 
	A          & 1/2 &  1 & 0 & 0 & 0 & 0 & -2  \\
	\tilde{A}  & 1/2 & -1 & 0 & 0 & 0 & 0 &  2  \\
	\hline
	\toprule
	\textbf{Theory $\mathcal{B}$} \ & U(1)_R & U(1)_B & U(1)_2 & U(1)_3 & U(1)_4  & \dots & U(1)_{n_g+1}\\
	\midrule
	X_{12}      & 1/2 &  1 & 0  & 0 & 0 & \dots & 2\nc+1-2n_g \\
	X_{21}      & 1/2 & -1 & 0  & 0 & 0 & \dots & -(2\nc+1-2n_g)  \\
	\vdots      & \vdots & \vdots  & \vdots & \vdots  & \vdots & \vdots & \vdots\\ 
	X_{n_g-2, n_g-1} & 1/2 & 1  & 0 &  -(2\nc-3)  & 2\nc-5 &  \dots & 0 \\
	X_{n_g-1,n_g-2}  & 1/2 & -1 &  0 &  2\nc-3  & -(2\nc-5) & \dots & 0 \\
	X_{n_g-1,n_g}   & 1/2 &  1 & -(2\nc-1) & 2\nc-3 &  0   & \dots & 0 \\
	X_{n_g,n_g-1}   & 1/2 &  -1 & 2\nc-1 & -(2\nc-3) &  0   & \dots & 0 \\
	X_{n_g,n_g+1}  & 1/2 &  1 & 2\nc-1 & 0 & 0   & \dots & 0 \\
	X_{n_g+1,n_g}  & 1/2 &  -1 & -(2\nc-1) & 0 & 0   & \dots & 0 \\
	S          & 1/2 &  1 & 0 & 0 & 0 & 0 & 0  \\
	\tilde{S}  & 1/2 & -1 & 0 & 0 & 0 & 0&  0  \\
	\bottomrule
\end{array}
\]
\caption{Baryonic symmetry map across the duality between a quiver of type $\mathcal{A}$ \eqref{eq:N=1_SO_quiver} and a quiver of type $\mathcal{B}$ \eqref{eq:N=1_USp_quiver}.}
\label{tab:baryonic}
\end{table}

We can therefore extend our S-duality claim as follows. Two quiver theories of type \eqref{eq:N=1_SO_quiver} and \eqref{eq:N=1_USp_quiver} are \textit{conformally dual} for fixed values of $N_c$ and $n_g$, and arbitrary number of adjoints and masses of the adjoints, as long as they are $\nn=1$ supersymmetric. Exact checks of this S-duality are the matching of the central charges and a nontrivial mapping of the global symmetries under which all~'t Hooft anomalies match. 

\subsection{\texorpdfstring{Relation to $\nn=2$ quiver theories}{Relation to N=2 quiver theories}}

We have performed the same checks for the $\mathcal{N}=2$ dualities originally studied in \cite{Uranga:1998uj}, finding a match for all the anomalies. The global symmetry map is the same as the one in the $\mathcal{N}=1$ case with $U(1)_B$ and $U(1)_R$ replaced by the following symmetry
\begin{equation}
	\begin{array}{c|cc}
		\toprule
		\  & SU(2)_R^{\mathcal{N}=2} & U(1)_R^{\mathcal{N}=2}\\
		\midrule
		T &  \multirow{2}{*}{$\square$} & 0 \\
		\widetilde{T}^{\dagger} & & 0 \\
		\hline
		X_{i,i-1} & \multirow{2}{*}{$\square$} & 0 \\
		X_{i-1,i}^{\dagger} & & 0 \\
		\hline
		\phi_i & 0 & 2 \\
	\end{array}
\end{equation}
where $T$ and $\widetilde{T}$ are either the symmetric and symmetric conjugate in theory $\mathcal{B}$ or the antisymmetric and antisymmetric conjugate in theory $\mathcal{A}$.
The central charges for the $\nn=2$ quivers are:
\begin{equation}
	a_{\mathcal{N}=2} = c_{\mathcal{N}=2} = \nc^2 \left(n_g+\frac{1}{2}\right)+\frac{1}{4} \nc \left(1-4 n_g^2\right)+\frac{1}{3} \left(n_g^3-n_g\right)\ .
\end{equation}
We note in passing that the $\mathcal{N}=1$ and $\mathcal{N}=2$ central charges satisfy $a_{\mathcal{N}=1} = 27/32 \ a_{\mathcal{N}=2}$ which matches the general result of \cite{Tachikawa:2009tt}.

\subsection{A further check from stringy instantons}
\label{stringy}
In this section we discuss a limiting case of the duality 
\begin{equation}
SO(2N_c+1) \times SU(2N_c-1)  \longleftrightarrow SU(2N_c) \times USp(2N_c-2)
\end{equation}
between model $\mathcal{A}$ in \eqref{eq:N=1_SO_quiver} and model $\mathcal{B}$ in \eqref{eq:N=1_USp_quiver} (respectively) with $n_g=1$.  The limiting case corresponds to formally taking $N_c=1$ (remember that we had $N_c>n_g$), i.e.  to the duality 
\begin{equation}\label{limiting}
SO(3)\times SU(1) \longleftrightarrow SU(2) \times USp(0)\ .
\end{equation}
On both sides we are in presence of a gauge node without any dynamical degrees of freedom but where it is still possible to have a nontrivial contribution at the nonperturbative level due to the presence of
Euclidean D$(-1)$-branes (a.k.a. E1-branes),  giving rise to an instanton contribution to the action. (See 
\cite{Argurio:2007vqa,Aharony:2007pr,Garcia-Etxebarria:2007fvo,Petersson:2007sc,Kachru:2008wt,Argurio:2008jm,Amariti:2008xu,Forcella:2008au,Bianchi:2012ud}
for the generation of nonperturbative superpotentials from stringy instantons in $\mathcal{N}=1$ quiver gauge theories.)
By integrating over the instantonic zero modes one is indeed left with an integral over Grassmann variables of the 
instanton action that gives rise to a non-vanishing superpotential for the fields involved in the model.
Here we match these nonperturbative contributions providing a nontrivial check of the duality in \eqref{limiting}.

\paragraph{$USp$ side.}

Let us focus first on the $SU(2)\times USp(0)$ side: here the $SU(2)$ gauge node has a pair of symmetric and conjugate symmetric 
representations that are equivalent to two adjoints. The relevant contribution of the `$O(1)$ instanton' in this case is
\begin{equation}
S_\text{inst}  \subset \sum_{a=1}^{2} \mu_a \epsilon_{ij} \phi_i \phi_j \mu_a^T\ ,
\end{equation}
where $\phi_{1,2}$ are the two adjoints and the action preserves the emergent $SU(2)$ global symmetry that arises if we consider
the gauge node $SU(2N_c-2)$ with $N_c=1$ in presence of a conjugate symmetric pair.
By integrating over the variables $\mu_i$ we arrive at the superpotential 
\begin{equation}
\label{stringyele}
W \propto \int e^{S_\text{inst}} d \mu_1 d\mu_2 \propto \text{Pf}^{\,2} [\phi_1,\phi_2] = \det [\phi_1,\phi_2] \propto \text{Tr} [\phi_1,\phi_2]^2\ .
\end{equation}
We are left with an $SU(2)$ gauge theory with two adjoints and the superpotential \eqref{stringyele},   that corresponds to the Leigh--Strassler (LS) fixed point for the $SU(2)$ gauge theory.

\paragraph{$SO$ side.}
In the second case we have $SO(2N_c+1)\times SU(2N_c-1)$ gauge group and there are two fields, that we denote $\phi_1$ and $\phi_2$,
that connect the two nodes and are charged under  $SO(2N_c+1)$ in the vectorial representations.
Again these fields are adjoints if $N_c=1$ and in this case the relevant contribution of the `$U(1)$ instanton' is
\begin{equation}
S_\text{inst}  \subset  \mu_1 \epsilon_{ij} \phi_i \phi_j \mu_2\ ,
\end{equation}
where again we wrote a manifestly symmetric action under the emergent $SU(2)$ flavor symmetry that exchanges the adjoints.
By integrating over the variables $\mu_i$ we arrive at the superpotential 
\begin{equation}
\label{stringymag}
W \propto \int e^{S_\text{inst}} d \mu_1 d\mu_2 \propto   \det [\phi_1,\phi_2] \propto \text{Tr} [\phi_1,\phi_2]^2\ .
\end{equation}
In this case we are left with an $SO(3)$ gauge theory with two adjoints and the superpotential \eqref{stringymag},  
that corresponds to the LS fixed point for the $SO(3)$ gauge theory.\\

All in all we have shown that the limiting case $N_c=1$ of the models in \eqref{limiting} corresponds to the duality between two $\mathcal{N}=1^*$ theories,
one with gauge group $SU(2)$ and the other with gauge group $SO(3) = SU(2)/\mathbb{Z}_2$, which is nothing but
S-duality.

%
%
%
\section{Explicit examples}
\label{sec:exam}
%
%
%

In this section we present some explicit examples of S-dual quivers considered in this paper. We briefly discuss the corresponding elliptic model and compute the central charges and the global symmetry map.  For low ranks we also compute the superconformal index in an expansion for small R-symmetry fugacities, providing another nontrivial check of the proposed duality.  Concretely, the 4d $\mathcal{N}=1$ superconformal index \cite{Kinney:2005ej,Romelsberger:2005eg} is defined as the following trace (see \cite{Gadde:2020yah,Rastelli:2016tbz} for recent reviews):
\begin{equation}
\mathcal{I} \equiv \Tr\, (-1)^{F} e^{-\beta  H_{S^3 \times S^1}} p^{J_1+\frac{R}{2}} q^{J_2 + \frac{R}{2}}
\prod_{b=1}^{\rk_F} v_b^{q_b} \ ,
\end{equation}
where $(-1)^F$ is the fermion number operator, $J_i$ are the angular momenta on the three-sphere, $R$ is the R-charge, and $q_b$ (not to be confused with $q$) are the conserved charges commuting with the supercharge,  the index $b$ running over the Cartan subgroup of the flavor symmetry group $F$, $b=1,\dots, \rk_F$. The quantities $p, q$ and $v_b$ are the associated fugacities.

For a gauge theory,  the index takes the form
\begin{equation}
\label{indexgen}
\mathcal{I} = \frac{(p;p)_\infty^{\rk_G} (q;q)_\infty^{\rk_G}}{|\text{Weyl}(G)|}
\oint_{T^{\rk_G}}  \prod_{i=1}^{\rk_G} \frac{dz_i}{2 \pi i z_i} 
\frac
{\prod_{I=1}^{n_\chi}
\prod_{\rho_I} 
\Gamma_e((pq)^{{R_I}/{2}} z^{\rho_I} v^{\omega _I};p,q)
}
{\prod_\alpha \Gamma_e(z^\alpha;p,q)}\ ,
\end{equation}
where $\rho_I$ ($\omega_I$) runs over the weight vectors of the representation $\mathcal{R}_I$ ($\mathcal{F}_I$) of the gauge (flavor) group of the $I$-th $\mathcal{N}=1$ matter multiplet ($n_\chi$ being their number), and $\alpha$ runs over the simple roots of the gauge group with rank $\rk G$. The holonomies $z_i$ are defined on the unit circle,  and the index $i$ runs over the Cartan subalgebra, $i=1,\dots, \rk_G$.  The quantities $(a;b)_\infty$ are $q$-Pochhammer symbols,
$
(a;b)_\infty \equiv \prod_{k=0}^\infty (1-a b^k)
$,
and $\Gamma_e$ are elliptic Gamma functions,
\begin{equation}
\Gamma_e(z;p,q)  \equiv \prod_{j,k=0}^\infty 
\frac{1-p^{j+1}q^{k+1} /z}{1-p^j q^k z}\ .
\end{equation}
In general, the match of the superconformal index for an $\mathcal{N}=1$ duality inherited via mass deformation of
an $\mathcal{N}=2$ duality can be derived using the formulation explained in \cite{Gadde:2010en}.
Nevertheless we will perform explicit checks in an expansion for small R-symmetry fugacities because we are not aware of the exact integral identities relating the indices of the $\mathcal{N}=2$ cases.

\subsection{\texorpdfstring{$\mathcal{N}=4$ $n_g=0$}{N=4 ng=0}}

\begin{figure}
\centering
\subfloat[IIA picture          \label{fig:N4ng0IIA}]
   {\includegraphics[width=.5\textwidth]{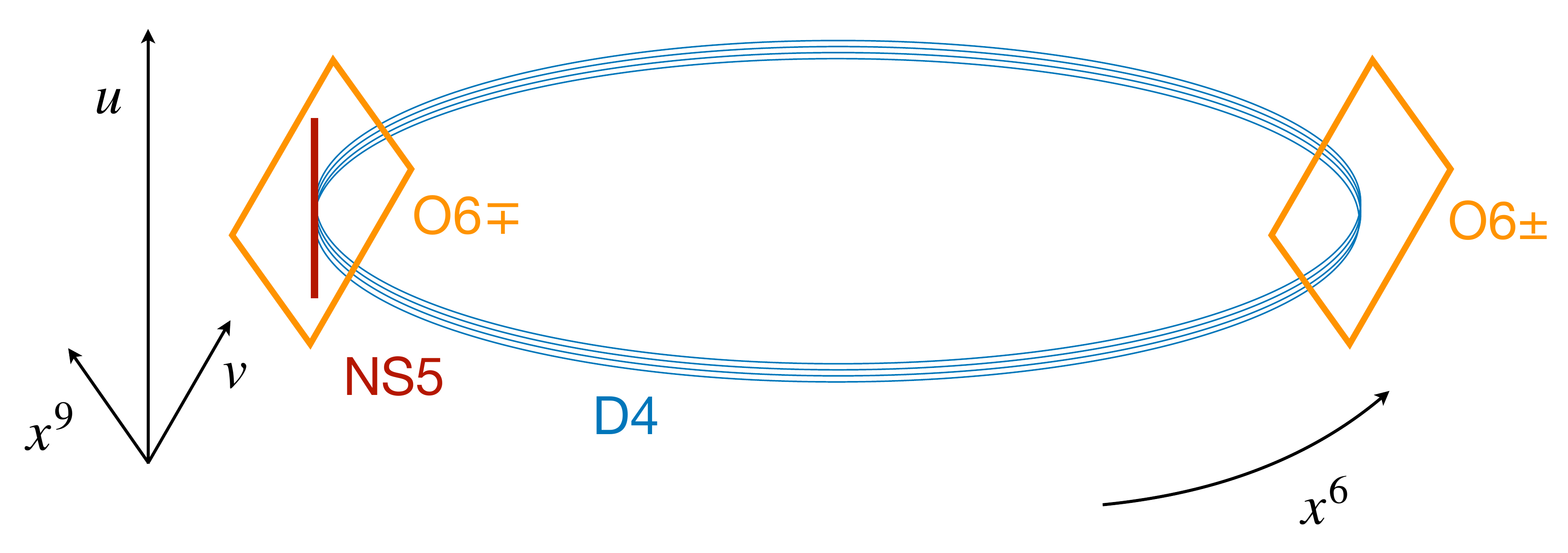}} \quad\quad
\subfloat[Quiver diagrams          \label{fig:N4ngquiv}]
   {\includegraphics[width=.4\textwidth]{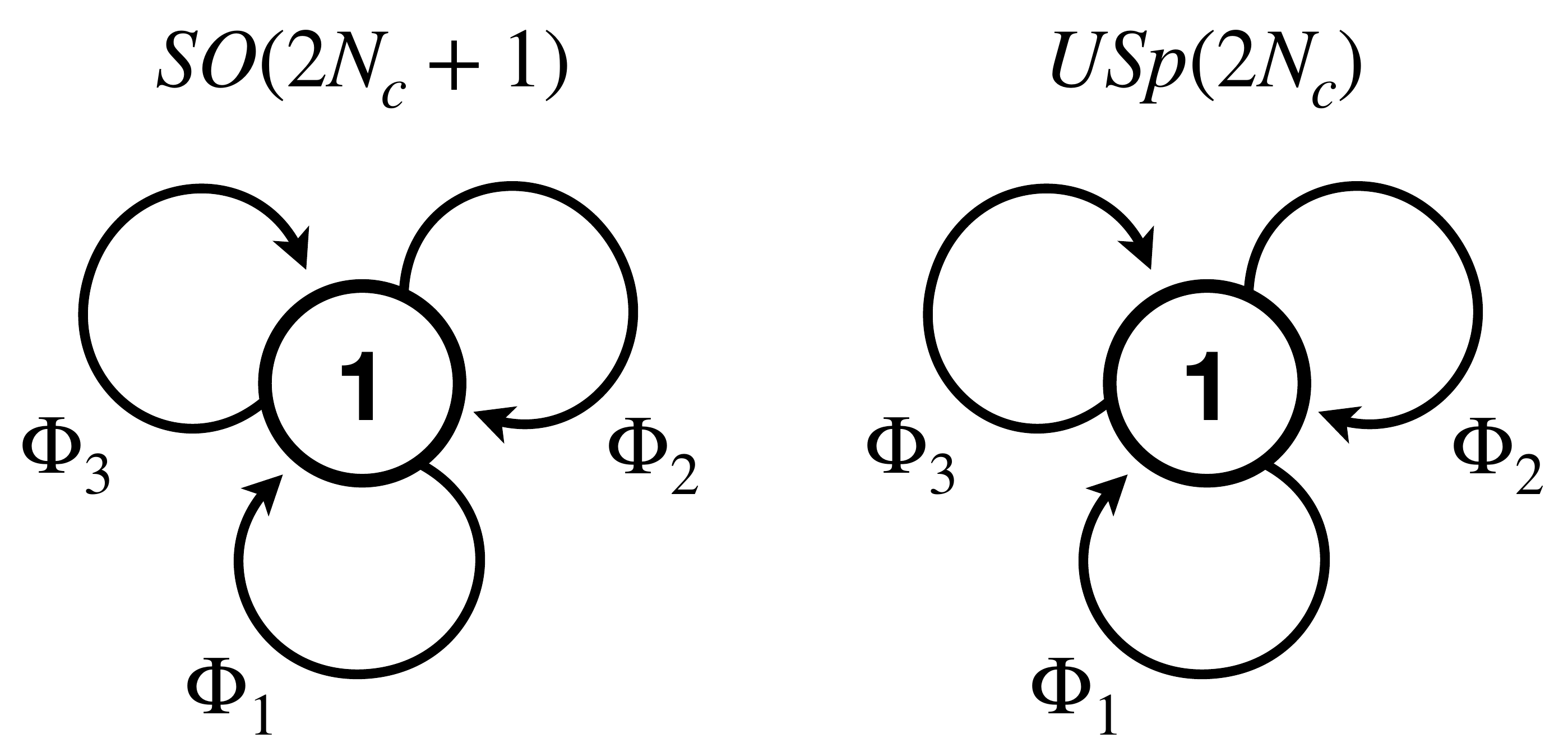}} 
         \caption{$\mathcal{N}=4$ S-dual quivers with $n_g=0$.}
         \label{fig:N4ng0}
\end{figure}

We start our analysis with the simplest case, corresponding to  $n_g=0$, i.e. the model with a single NS5-brane (see figure \ref{fig:N4ng0IIA}) that exhibits an $\mathcal{N}=4$ supersymmetry. 
If we place an O6$^-$-plane (O6$^+$) on the top of the NS5-brane and an O6$^+$ (O6$^-$) on the top of the D4-branes, we obtain an $\mathcal{N}=4$ gauge theory with an $SO(2N_c+1)$ ($USp(2N_c)$) gauge group and three adjoint fields $\phi_1$,  $\phi_2$,  $\phi_3$. All these fields combine to form an $\mathcal{N}=4$ vector multiplet. %
The S-duality between these theories has  been studied in \cite{Dorey:1996hx,Argyres:2006qr}.

\subsection{\texorpdfstring{$\mathcal{N}=1$ $n_g=0$}{N=1 ng=0}}
\begin{figure}
  \centering
\subfloat[IIA picture \label{fig:N1ng0IIA}]
   {\includegraphics[width=.5\textwidth]{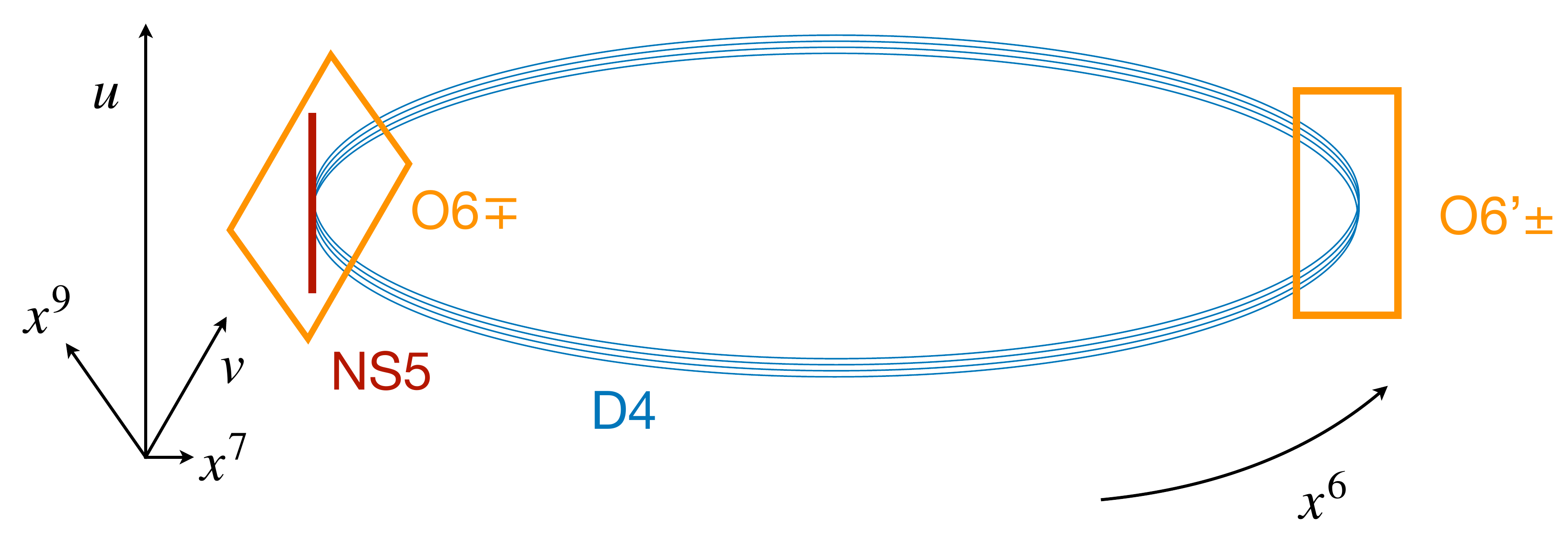}} \quad\quad
\subfloat[Quiver diagrams \label{fig:N1ng0quiv}]
{\includegraphics[width=.4\textwidth]{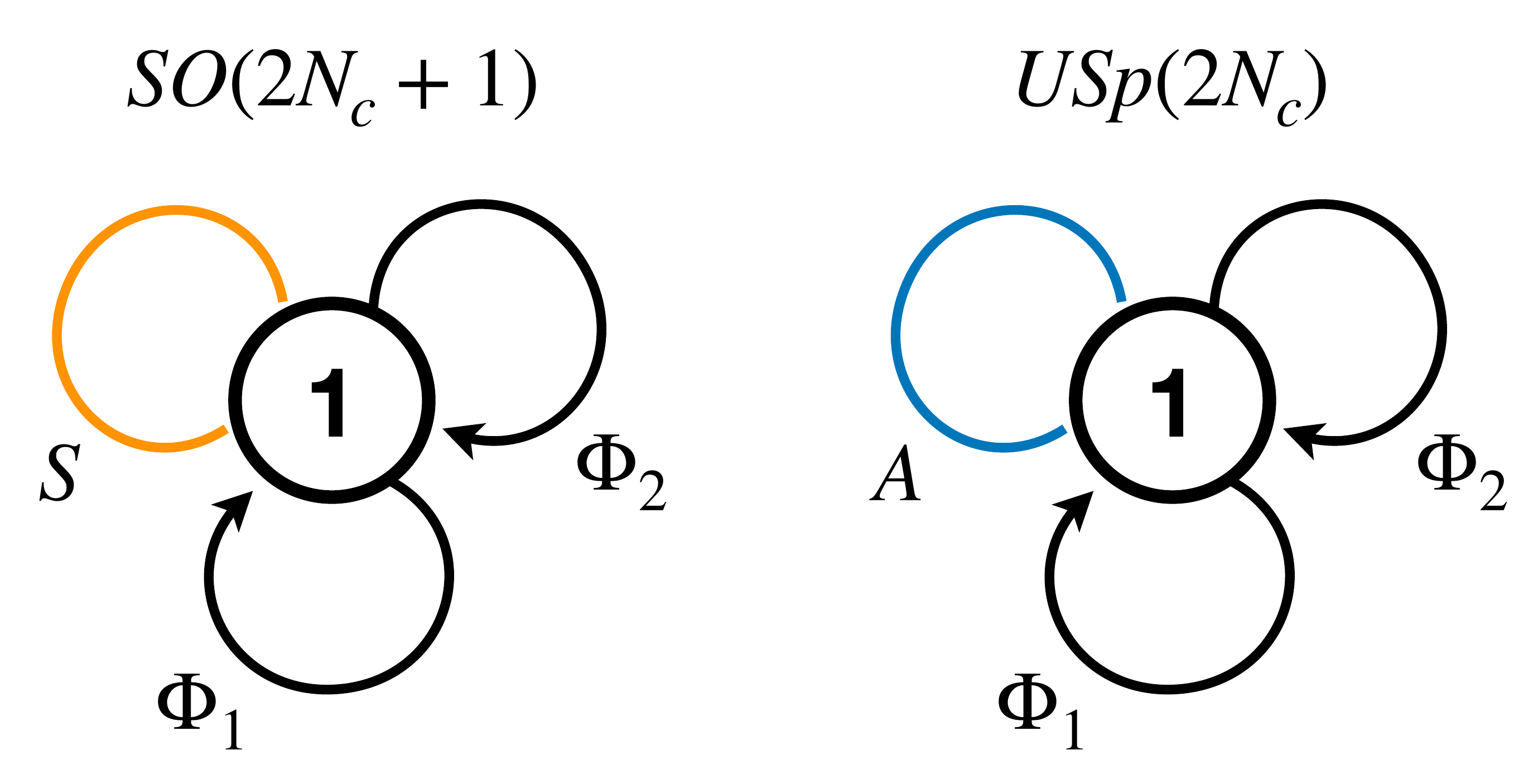}}
\caption{$\mathcal{N}=1$ S-dual quivers with $n_g=0$.}
\label{fig:N1ng0}
\end{figure}

Starting from the models in figure \ref{fig:N4ng0}, we can break part of the supersymmetry by tilting the O6-plane on the top of the D4-branes. We obtain the $\mathcal{N}=1$ S-dual models depicted in figure \ref{fig:N1ng0}. If the RR charges of the O-planes are O6$^-$-O6'$^+$ (O6$^+$-O6'$^-$) the resulting quiver has an $SO(2N_c+1)$ ($USp(2N_c)$) gauge group, with two fields $\phi_1, \ \phi_2$ in the adjoint representation and a field in the symmetric (antisymmetric) representation.  %
The superpotentials of the theories are
\begin{equation}
  W_{\mathcal{A}} = \phi_1 \phi_2 S\ , \qquad W_{\mathcal{B}} = \phi_1 \phi_2 A\ .
  \label{N1ng0:W}
\end{equation}
In both theories the R-charges of the fields are obtained by imposing the vanishing of the $\beta$ functions, and are given by
\begin{equation}
  R[\phi_1] = R[\phi_2] = \frac{1}{2}\ ,  \qquad R[S] = R[A] = 1\ .
  \label{N1ng0:R}
\end{equation}
We claim that these models are \textit{conformally dual} to $\mathcal{N}=1^*$  LS fixed points with
$SO(2N_c+1)$ and $USp(2N_c)$ gauge group respectively (see \cite{Kumar:2001iu,Kim:2004xx,Wyllard:2007qw} for discussions of this duality).
 It can be checked indeed that they have the same central charges and superconformal indices 
of the  $\mathcal{N}=1^*$  fixed points.
Furthermore there is a quadratic exactly marginal deformation in both cases, involving the field $S$ and $A$ respectively, that can be added to the
two models, implying a motion on the conformal manifold.
By adding such a deformation and \emph{integrating out} the fields $S$ and $A$ the duality reduces to the duality between two LS fixed points.

\subsection{\texorpdfstring{$\mathcal{N}=2$ $n_g=1$}{N=2 ng=1}}
\begin{figure}
	\centering
	\subfloat[IIA picture \label{fig:N2ng1IIA}]
	{\includegraphics[width=.45\textwidth]{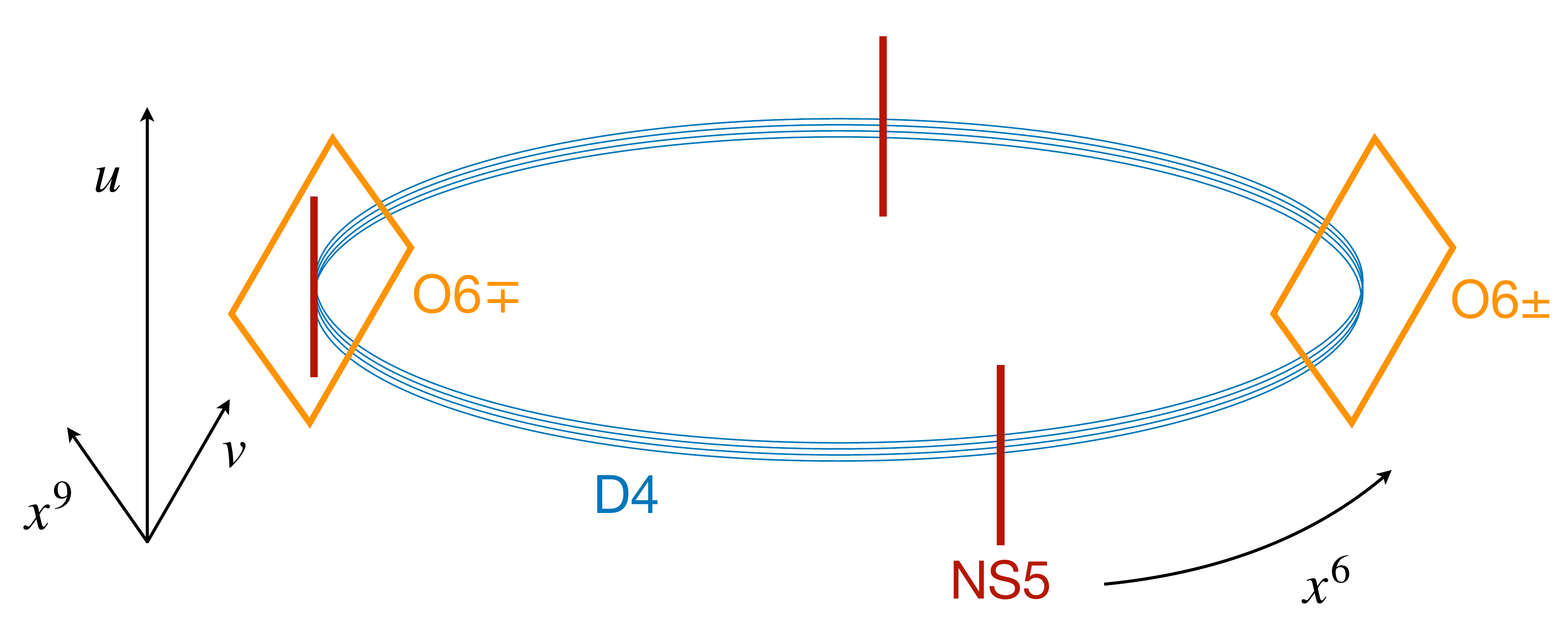}}
	 \quad\quad
	\subfloat[Quiver diagrams \label{fig:N2ng1quiv}]
	{\includegraphics[width=.45\textwidth]{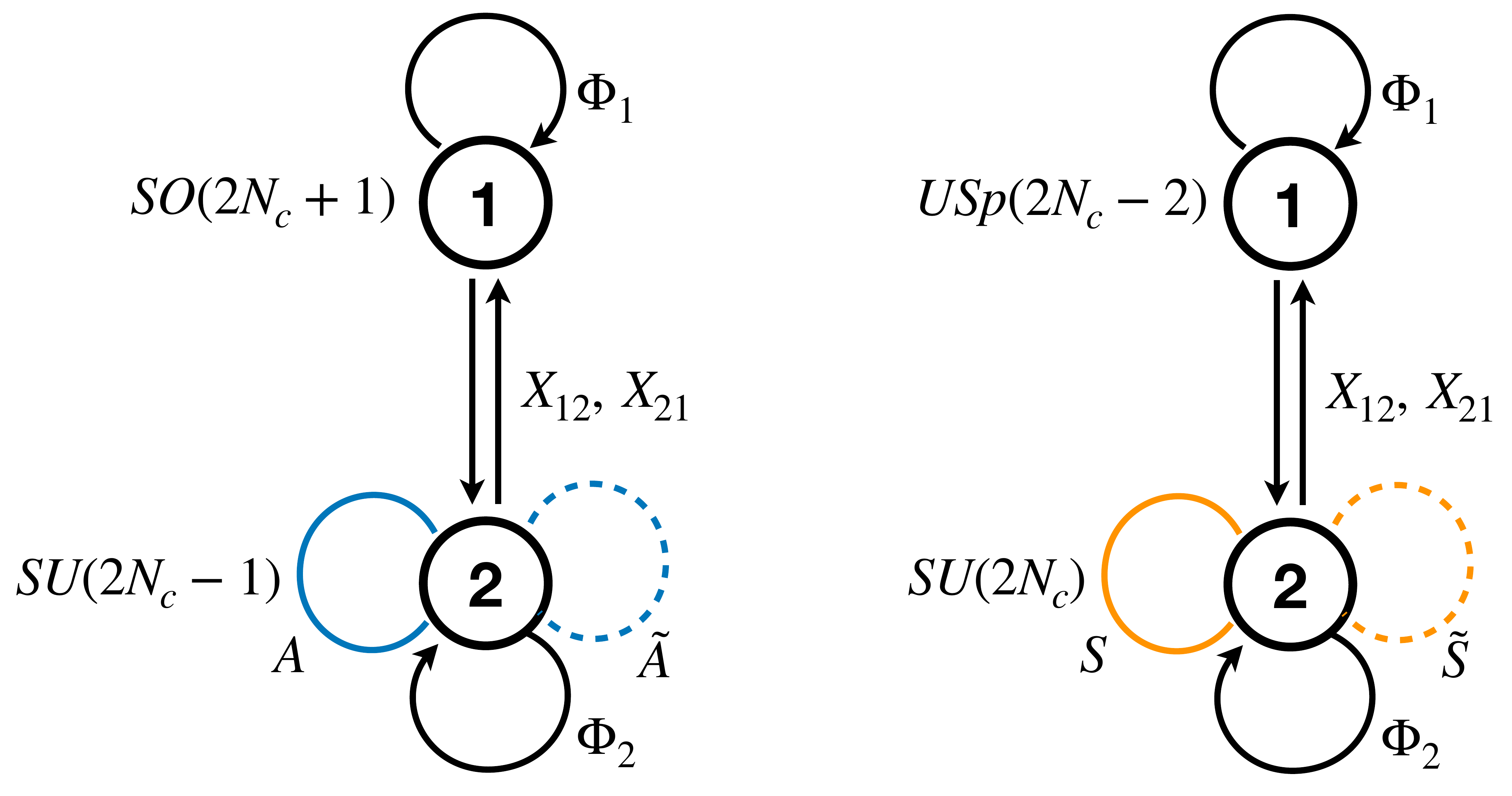}}
	\caption{$\mathcal{N}=2$ S-dual quivers with $n_g=1$.}
	\label{fig:N2ng1}
\end{figure}

We begin our analysis with two S-dual $\mathcal{N}=2$ theories with $n_g=1$ (see figure \ref{fig:N2ng1quiv}).  Both can be obtained from the same elliptic model (given by three parallel NS5-branes with three stacks of D4-branes suspended between them) by placing an O6-plane on top of an NS5-brane and the other on a stack of D4-branes (see figure \ref{fig:N2ng1IIA}). The left picture is realized when the O-planes are chosen with RR charges O6$^\pm$ respectively, while the right picture by choosing O6$^\mp$.

\paragraph{Model $\mathcal{A}$ \eqref{eq:N=2_SO_quiver}.}
In the first quiver the fields $\phi_1$ and $\phi_2$ are in the adjoint representations of $SO(2N_c+1)$ and $SU(2N_c-1)$ (respectively,) the fields $X_{12},X_{21}$ are in the fundamental, anti-fundamental representation (respectively) of $SU$ and in the vectorial representation of $SO$, while the fields $A, \tilde{A}$ are in the antisymmetric and conjugate antisymmetric representations (respectively) of $SU$. All these fields combine in $\mathcal{N}=2$ hypermultiplets and vector multiplets. The superpotential of the theory is given by
\begin{equation}
	W_{ \mathcal{A}}= \phi_1 X_{12} X_{21} - \phi_2 X_{21}X_{12} + \phi_2 A \tilde{A}\ ,
	\label{N2ng1:WSO}
\end{equation}
The ranks of the gauge groups are obtained by imposing the vanishing of the $\beta$-functions, which is equivalent to require the anomaly freedom of the R-symmetry:
\begin{align}
	&(2N_c-1) (1+R[\phi_1]-1) + (2N_c-1)(R[X_{12}]-1+R[X_{21}]-1)=0\ , \\
	\begin{split}
		&(2N_c-1) (1+R[\phi_2]-1) + \frac{1}{2} (2N_c+1)(R[X_{12}]-1+R[X_{21}]-1)\ +\\
		&+\frac{(2N-1-2)}{2} (R[A]-1+R[\tilde{A}]-1) =0\ .
	\end{split}
\end{align}
The central charges of the theory are
\begin{equation}
	a_{ \mathcal{A}} = c_{ \mathcal{A}} = \frac{3}{4}N_c(2N_c-1)\ .
	\label{N2ng1:acSO}
\end{equation}
Let us call $b$ the fugacity of $SU(2)_R$ (which is a flavor symmetry from the $\mathcal{N}=1$ perspective) and $t$ that of $U(1)_2$ in table \ref{N2ng1:tab}. Then in the smallest case of $N_c =2$ the superconformal index reads 
\begin{align}
\mathcal{I} = &\  1+ 2 (p q)^{2/3} (1+b^2) -2 (p+q) (p q)^{1/3}+ 3 pq (b^2-1) + (p+q) (p q)^{2/3} (3+b^2)  + \nonumber \\
& -2 \left(p^2+q^2\right) (p q)^{1/3}+  (p q)^{4/3} \left((3 b^3-b) t^6+\frac{ (3 b^3-b)}{t^6}+\frac{8 b^6+7 b^4+2}{b^2}-4\right) + \nonumber \\
& - pq (p+q) \,\frac{2 b^3 \left(t^{12}+1\right)+5 b^2 t^6+7 t^6}{t^6} + \ldots\ , \label{eq:SCIN=2n=1}
\end{align}
whereas for $N_c=3$ it reads
\begin{align}
\mathcal{I} = &\  1+ 2 (p q)^{2/3} (1+b^2) -2 (p+q) (p q)^{1/3}+ 3 pq (b^2-1) + (p+q) (p q)^{2/3} (3+b^2)  + \nonumber \\
& -2 \left(p^2+q^2\right) (p q)^{1/3}+  (p q)^{4/3} \left( \left(2 b^3-b\right) t^{10}+\frac{\left(2 b^3-b\right)}{t^{10}}+\frac{7 b^6+6 b^4+2}{b^2}-3   \right) + \nonumber \\
& - pq (p+q) \, \frac{b^3 t^{20}+b^3+4 \left(b^2+2\right) t^{10}}{t^{10}}+ \ldots\ . \label{eq:SCIN=2n=2}
\end{align}

\paragraph{Model  $\mathcal{B}$ \eqref{eq:N=2_USp_quiver}.}
In the second quiver, the fields $\phi_1$ and $\phi_2$ are in the adjoint representations of $USp(2N_c-2)$ and $SU(2N_c)$ (respectively), the fields $X_{12},X_{21}$ are in the fundamental, anti-fundamental representation (respectively) of $SU$ and in the fundamental  representation of $USp$, while the fields $S, \tilde{S}$ are in the symmetric and conjugate symmetric representations (respectively) of $SU$. The superpotential of the theory is
\begin{equation}
	W_{ \mathcal{B}}= \phi_1 X_{12} X_{21} - \phi_2 X_{21}X_{12} + \phi_2 S \tilde{S}\ .
	\label{N2ng1:WUSp}
\end{equation}
As in theory $\mathcal{A}$, every matter fields has R-charge $2/3$ and the ranks of the gauge groups are determined using the $\beta$-functions. The central charges are given by
\begin{equation}
	a_{ \mathcal{B}} = c_{ \mathcal{B}} = \frac{3}{4}N_c(2N_c-1)\ ,
	\label{N2ng1:acUSp}
\end{equation}
and they match with the dual case, as expected.

The matter content and the charges of both theories are summarized in table \ref{N2ng1:tab}. In addition to the usual $SU(2)_R \times U(1)_R$ R-symmetry, there is an extra $U(1)_2$ global symmetry. We have checked that given the charge assignment in table \ref{N2ng1:tab} 
the 't Hooft anomalies of the two theories match.  For $N_c=2,3$ the superconformal index matches with \eqref{eq:SCIN=2n=1} and \eqref{eq:SCIN=2n=2} respectively, as expected.

\paragraph{Gauge group global structure.} It is interesting to comment on the global properties of the gauge group \cite{Aharony:2013hda,Gaiotto:2014kfa}. 
So far we have not distinguished (by abuse of notation) the $SO(2N_c+1)$ gauge group from $Spin(2N_c+1)$. Indeed the different global structure does not affect the central charges, the 't Hooft anomalies or the superconformal index.  Nevertheless the dual pairs discussed here have nontrivial choices of
global structure and it is useful to study their behavior under the dualities we are discussing.\footnote{An observable which is sensitive to the global structure of the gauge group is e.g. the Lens index. It would be possible to compute and match such index across dual phases along the lines of \cite{Amariti:2019but,Razamat:2013opa}. We are grateful to the referee for suggesting this possibility.} 
For example  in the case with $n_g=1$ and $N_c=0$  we have already observed in section \ref{stringy} that the action of S-duality explicitly maps a model with $Spin(3)/\mathbb{Z}_2$ gauge group to a model with $USp(2)$ gauge group. For higher $N_c$ a similar relation holds. In this case we 
can look at the structure of the lattices of charges of mutually local bound states of Wilson and t 'Hooft lines. These lattices can be organized in terms of the
subgroup of the central symmetry unbroken by the matter fields (see \cite{Amariti:2016hlj} for an analogous discussion in the case of elliptic models).  In this case the center symmetry is $\mathbb{Z}_2\times \mathbb{Z}_2$ and the matter fields break it to 
$\mathbb{Z}_2^{\text{diag}}$. 
Line operators can be organized in terms of the two electric charges (the charges of the Wilson lines under the center) and magnetic charges (the charges of the 't Hooft lines under the center), $(e_1,e_2;m_1,m_2)$, and pairs of lines must respect a condition of mutual locality imposed by the DSZ quantization condition. Using additivity and linearity one can show that $m_1=m_2$ and that only the sum of the electric lines is relevant. The generators of the lattices are then given by a pair of vectors $(e_a,m_a)$ and $(e_b,m_b)$.
There are three possible independent choices given by the sets $\{(1,0),(0,2)\}$, $\{(2,0),(0,1)\}$ and $\{(2,0),(1,1)\}$.  The first two possibilities are mapped under the electro-magnetic charge conjugation. For example when model $\mathcal{A}$ has gauge group $Spin(2N_c+1) \times SU(2N_c-1)$ the S-dual theory given by model $\mathcal{B}$  has gauge group $(USp(2N _c-2)\times SU(2N_c))/\mathbb{Z}_2$.
Vice-versa if model $\mathcal{A}$ has gauge group $USp(2 N_c-2)\times SU(2N_c)$ the S-dual theory given by model $\mathcal{B}$  has gauge group 
$(Spin(2N_c+1) \times SU(2N_c-1))/\mathbb{Z}_2$.
This discussion applies equivalently to the models with higher $n_g$, by observing that also in that case only a diagonal $\mathbb{Z}_2$ symmetry is unbroken once the matter fields are specified. It would be interesting to reproduce this discussion from the M-theory analysis of \cite{Albertini:2020mdx}.%
\begin{table}
	\renewcommand{\arraystretch}{1.5}
	\[
	\begin{array}{lccccc}
		\toprule
		\ & \quad SO(2N_c+1) \quad & \quad SU(2N_c-1) \quad & \quad SU(2)_R \quad & \quad U(1)_R \quad & \quad U(1)_2 \quad \\
		\midrule
		\phi_1 & {\bf Adj} & \mathbf{1} & 0 & 2 & 0 \\
		X_{12} & \tiny{\yng(1)} & \overline{\tiny{\yng(1)}} & \multirow{2}{*}{\tiny{\yng(1)}} & 0 & 2N_c-2 \\
		X_{21} & \tiny{\yng(1)} & \tiny{\yng(1)} & & 0 & -(2N_c-2) \\
		\phi_2 & \mathbf{1} & {\bf Adj} & 0 & 2 & 0 \\
		A & \mathbf{1} & \tiny{\yng(1,1)} & \multirow{2}{*}{\tiny{\yng(1)}} & 0 & -2 \\
		\tilde{A} & \mathbf{1} & \overline{\tiny{\yng(1,1)}} & & 0 & 2 \\
		\toprule
		\ & \quad USp(2N_c-2) \quad & \quad SU(2N_c) \quad & \quad SU(2)_R \quad & \quad U(1)_R \quad & \quad U(1)_2 \quad \\
		\midrule
		\phi_1 & {\bf Adj}& \mathbf{1} & 0 & 2 & 0 \\
		X_{12} & \tiny{\yng(1)} & \overline{\tiny{\yng(1)}} & \multirow{2}{*}{\tiny{\yng(1)}} & 0 & 2N_c-1 \\
		X_{21} & \tiny{\yng(1)} & \tiny{\yng(1)} & & 0 & -(2N_c-1) \\
		\phi_2 & \mathbf{1} & {\bf Adj} & 0 & 2 & 0 \\
		S & \mathbf{1} & \tiny{\yng(2)} & \multirow{2}{*}{\tiny{\yng(1)}} & 0 & 0 \\
		\tilde{S} & \mathbf{1} & \overline{\tiny{\yng(2)}} & & 0 & 0 \\
		\bottomrule
	\end{array}
	\]
	\caption{Matter content and charges of $\mathcal{N}=2$ S-dual quivers with $n_g=1$. Calling $J_3$ and $T$ the generators of the $\mathcal{N}=2$ R-symmetry $SU(2)_R \times U(1)_R$ (respectively), the generator of the $\mathcal{N}=1$ R-symmetry is given in our conventions by $\frac{2}{3}J_3 + \frac{1}{3}T$, so that all matter fields have $\mathcal{N}=1$ R-charge $2/3$.}  
	\label{N2ng1:tab}
\end{table}

\subsection{\texorpdfstring{$\mathcal{N}=2$ $n_g=2$}{N=2 ng=2}}
\begin{figure}[h]
	\centering
	\includegraphics[width=0.5\textwidth]{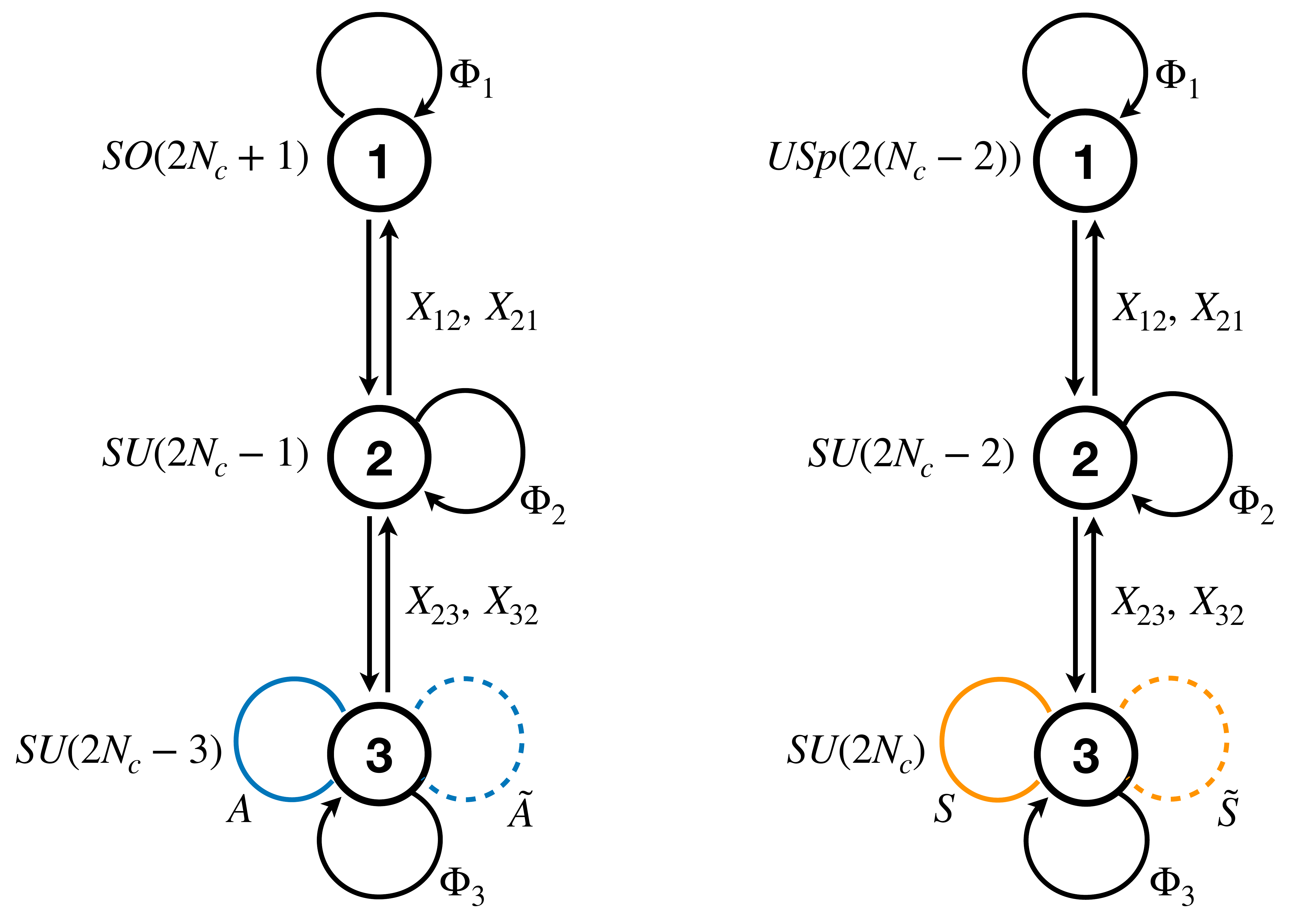}
	\caption{$\mathcal{N}=2$ S-dual quiver diagrams with $n_g=2$.}
	\label{fig:N2ng2}
\end{figure}

Adding a node to the previous quivers, using a rank assignment such that the  $\beta$ functions vanish, we obtain the two theories depicted in figure \ref{fig:N2ng2}, which are again S-dual.
The superpotentials of the theories are:
\begin{equation}
	\begin{split}
		& W_{ \mathcal{A}} = \phi_1 X_{12}X_{21} - \phi_2 X_{21}X_{12} + \phi_2 X_{23}X_{32} - \phi_3 X_{32}X_{23} + \phi_3 A \tilde{A}\ , \\
		& W_{ \mathcal{B}} = \phi_1 X_{12}X_{21} - \phi_2 X_{21}X_{12} + \phi_2 X_{23}X_{32} - \phi_3 X_{32}X_{23} + \phi_3 S \tilde{S}\ .
	\end{split}
	\label{N2ng2:W}
\end{equation}
Their matter content and charges are listed in table \ref{N2ng2:tab}, and their central charges are given by:
\begin{equation}
	a_{ \mathcal{A}} = c_{ \mathcal{A}} = a_{ \mathcal{B}} = c_{ \mathcal{B}} = 2+\frac{5}{4}N_c(2N_c-3)\ .
	\label{N2ng2:ac}
\end{equation}
In both theories, in the smallest case of $N_c=3$ the superconformal index reads
\begin{align}
\mathcal{I} = &\  1+ 3 (p q)^{2/3} (1+b^2) -3 (p+q) (p q)^{1/3}+ pq (5b^2-4) + (p+q) (p q)^{2/3} (4+b^2)  + \nonumber \\
& -3 \left(p^2+q^2\right) (p q)^{1/3}+  (p q)^{4/3} \left( \frac{11 \left(b^6+b^4\right)+3}{b^2}-5 \right) + \nonumber \\
& - 4pq (p+q)(3+2b^2) + \ldots\ ,
\end{align}
whereas for $N_c=4$ we were able to compute it only up to order $(p+q) (p q)^{1/3}$, and for both it reads
\begin{align}
\mathcal{I} =  1+ 3 (p q)^{2/3} (1+b^2) -3 (p+q) (p q)^{1/3} + \ldots\ .
\end{align}

\begin{table}
\begin{adjustbox}{width=\columnwidth,center}
	\renewcommand{\arraystretch}{1.5}
	\begin{tabular}{lccccccc}
		\toprule
		 & $SO(2N_c+1)$ &  $SU(2N_c-1)$  &  $SU(2N_c-3)$ &  $SU(2)_R$  &  $U(1)_R$  &  $U(1)_2$  &  $U(1)_3$ \\
		\midrule
		$\phi_1$ & $\mathbf{Adj}$ & $\mathbf{1}$ & $\mathbf{1}$ & 0 & 2 & 0 & 0 \\
		$X_{12}$ & $\tiny{\yng(1)}$ & $\overline{\tiny{\yng(1)}}$ & $\mathbf{1}$ & \multirow{2}{*}{$\tiny{\yng(1)}$ } & $0$ & $2N_c-2$ & 0 \\
		$X_{21}$ & $\tiny{\yng(1)}$ & $\tiny{\yng(1)}$ & $\mathbf{1}$ & & $0$ & $-(2N_c-2)$ & $0$ \\
		$\phi_2$ & $\mathbf{1}$ & $\mathbf{Adj}$ & $\mathbf{1}$ & $0$ & $2$ & $0$ & $0$ \\
		$X_{23}$ & $\mathbf{1}$ & $\tiny{\yng(1)}$ & $\overline{\tiny{\yng(1)}}$ & \multirow{2}{*}{$\tiny{\yng(1)}$ } & $0$ & $-(2N_c-2)$ & $2N_c-4$ \\
		$X_{32}$ & $\mathbf{1}$ & $\overline{\tiny{\yng(1)}}$ & $\tiny{\yng(1)}$ &  & 0 & $2N_c-2$ & $-(2N_c-4)$ \\
		$\phi_3$ & $\mathbf{1}$ & $\mathbf{1}$ & $\mathbf{Adj}$ & $0$ & $2$ & $0$ & $0$ \\
		$A$ & $\mathbf{1}$ & $\mathbf{1}$ & $\tiny{\yng(1,1)}$ & $\multirow{2}{*}{\tiny{\yng(1)}}$ & $0$ & $0$ & $-2$ \\
		$\tilde{A}$ & $\mathbf{1}$ & $\mathbf{1}$ & $\overline{\tiny{\yng(1,1)}}$ & & $0$ & $0$ & $2$ \\
		\toprule
		 & $USp(2N_c-4)$ &  $SU(2N_c-2)$ &  $SU(2N_c)$ &  $SU(2)_R$ &  $U(1)_R$  &  $U(1)_2$  &  $U(1)_3$ \\
		\midrule
		$\phi_1$ & $\mathbf{Adj}$ & $\mathbf{1}$ & $\mathbf{1}$ & $0$ & $2$ & $0$ & $0$ \\
		$X_{12}$ & $\tiny{\yng(1)}$ & $\overline{\tiny{\yng(1)}}$ & $\mathbf{1}$ & \multirow{2}{*}{$\tiny{\yng(1)}$} & $0$ & $-(2N_c-1)$ & $2N_c-3$ \\
		$X_{21}$ & $\tiny{\yng(1)}$ & $\tiny{\yng(1)}$ & $\mathbf{1}$ & & $0$ & $2N_c-1$ & $-(2N_c-3)$ \\
		$\phi_2$ & $\mathbf{1}$ & $\mathbf{Adj}$ & $\mathbf{1}$ & $0$ & $2$ & $0$ & $0$ \\
		$X_{23}$ & $\mathbf{1}$ & $\tiny{\yng(1)}$ & $\overline{\tiny{\yng(1)}}$ & \multirow{2}{*}{ $\tiny{\yng(1)}$ } & $0$ & $2N_c-1$ & 0 \\
		$X_{32}$ & $\mathbf{1}$ & $\overline{\tiny{\yng(1)}}$ & $\tiny{\yng(1)}$ &  & $0$ & $-(2N_c-1)$ & $0$ \\
		$\phi_3$ & $\mathbf{1}$ & $\mathbf{1}$ & $\mathbf{Adj}$ & $0$ & $2$ & $0$ & $0$ \\
		$S$ & $\mathbf{1}$ & $\mathbf{1}$ & $\tiny{\yng(2)}$ & \multirow{2}{*}{$\tiny{\yng(1)}$} & $0$ & $0$ & $0$ \\
		$\tilde{S}$ & $\mathbf{1}$ & $\mathbf{1}$ & $\overline{\tiny{\yng(2)}}$ & & $0$ & $0$ & $0$ \\ 
		\bottomrule
	\end{tabular}
\end{adjustbox}
	\caption{Matter content and charges of $\mathcal{N}=2$ S-dual quivers with $n_g=2$.}  
	\label{N2ng2:tab}
\end{table}

\subsection{\texorpdfstring{$\mathcal{N}=1$ $n_g=1$}{N=1 ng=1}}
\begin{figure}
  \captionsetup[subfigure]{labelformat=empty}
	\centering
	\subfloat[][Tilting two NS5-branes.]
	{\includegraphics[width=.5\textwidth]{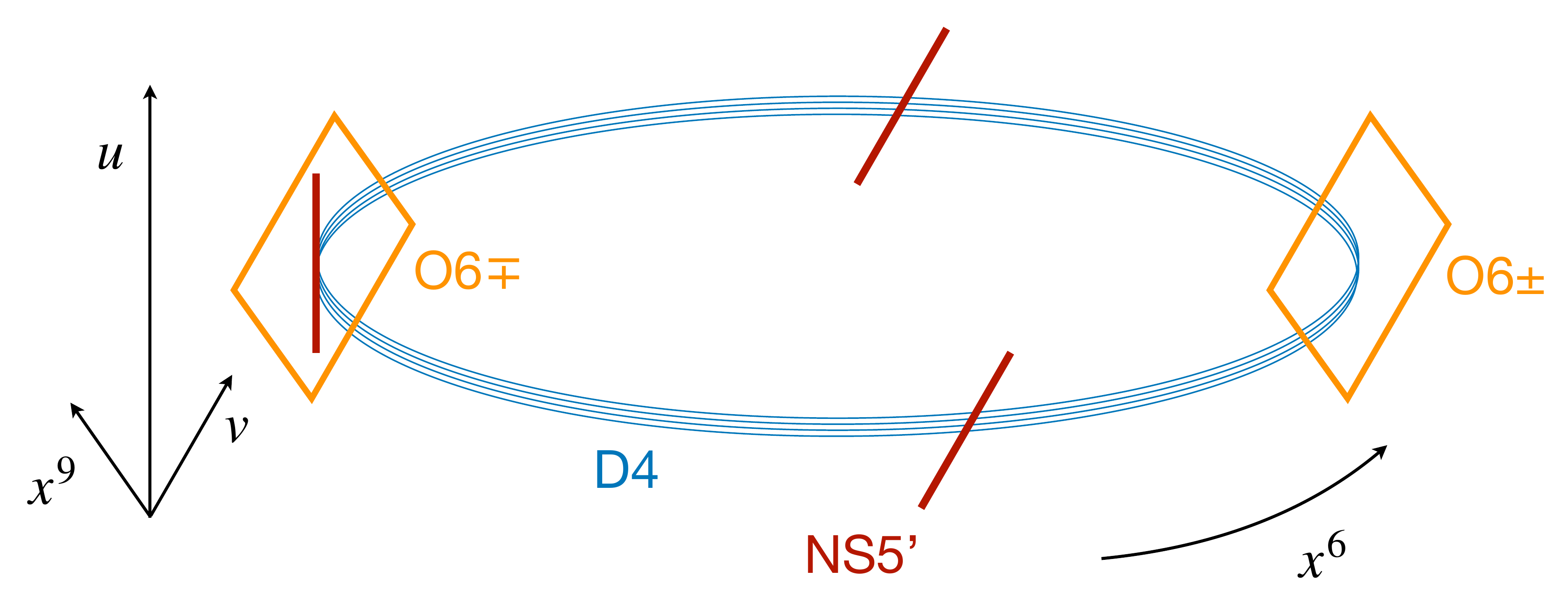}} \quad
	\subfloat
	{\includegraphics[width=.4\textwidth]{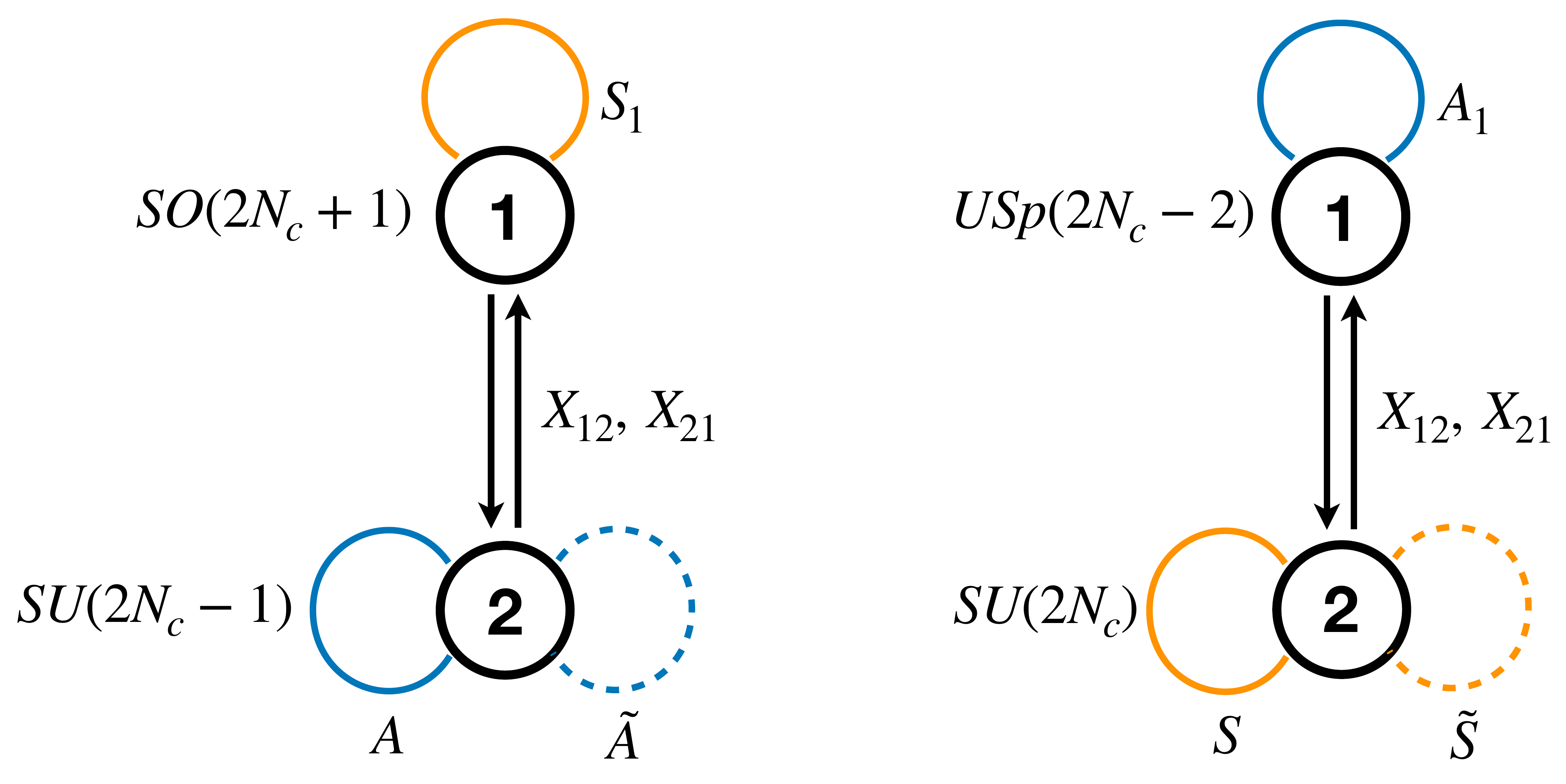}} \\
	\subfloat[][Tilting an O6-plane.]
	{\includegraphics[width=.5\textwidth]{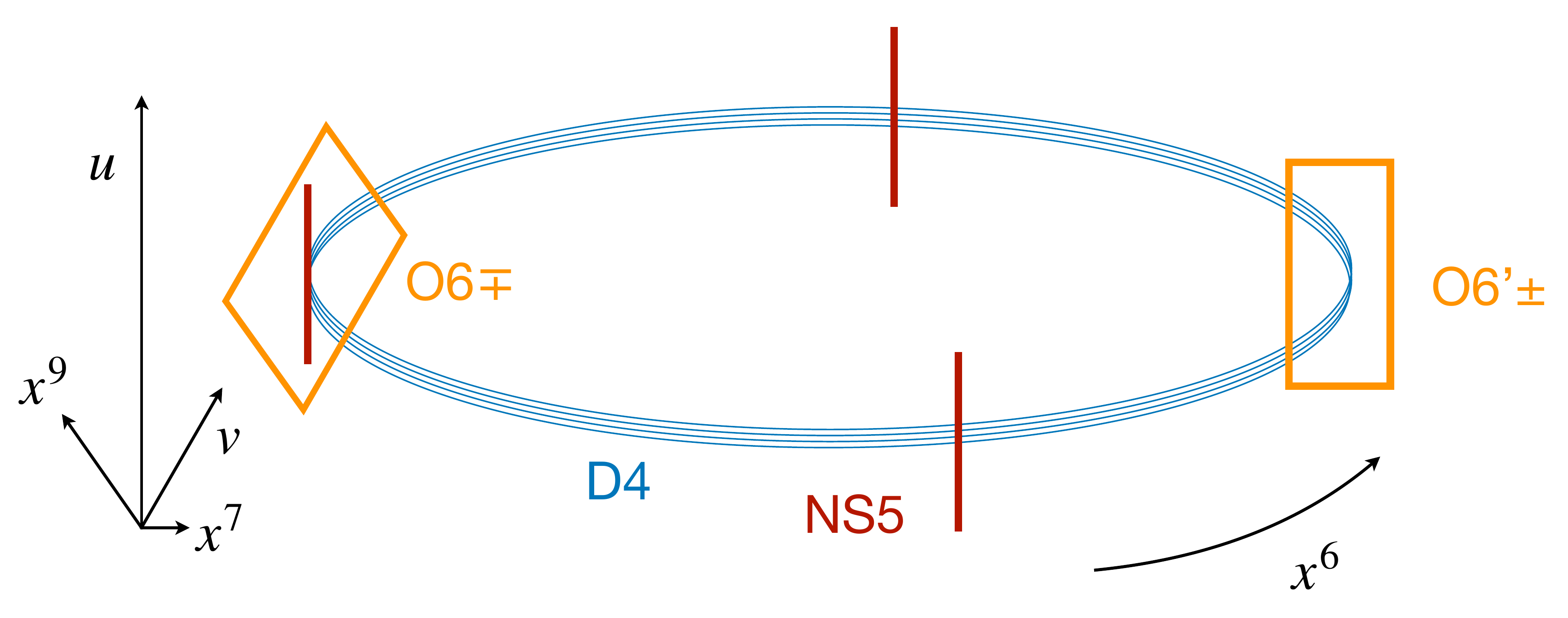}} \quad
	\subfloat
	{\includegraphics[width=.4\textwidth]{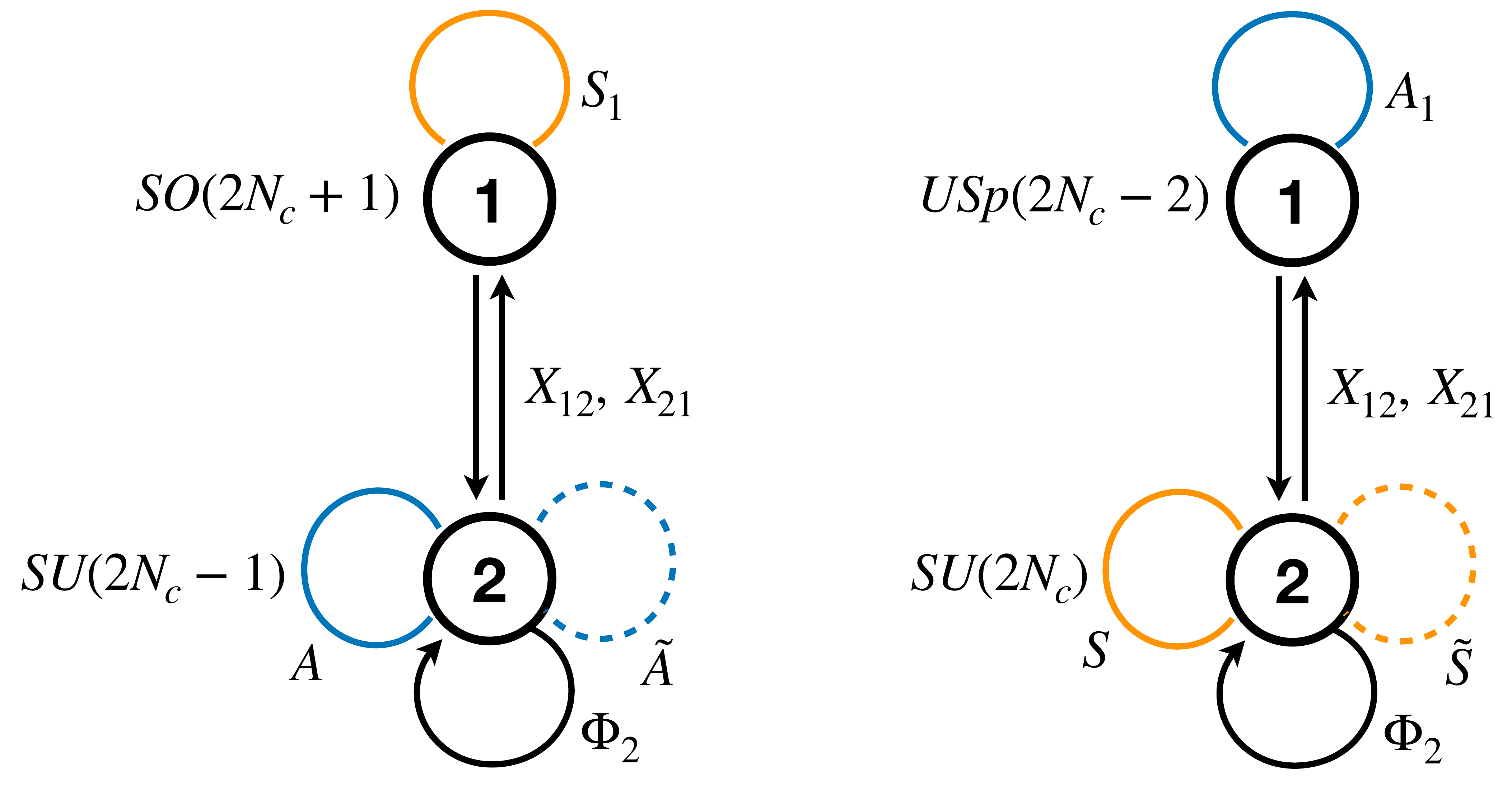}} \\
	\subfloat[][Tilting two NS5-branes and an O6-plane.]
	{\includegraphics[width=.5\textwidth]{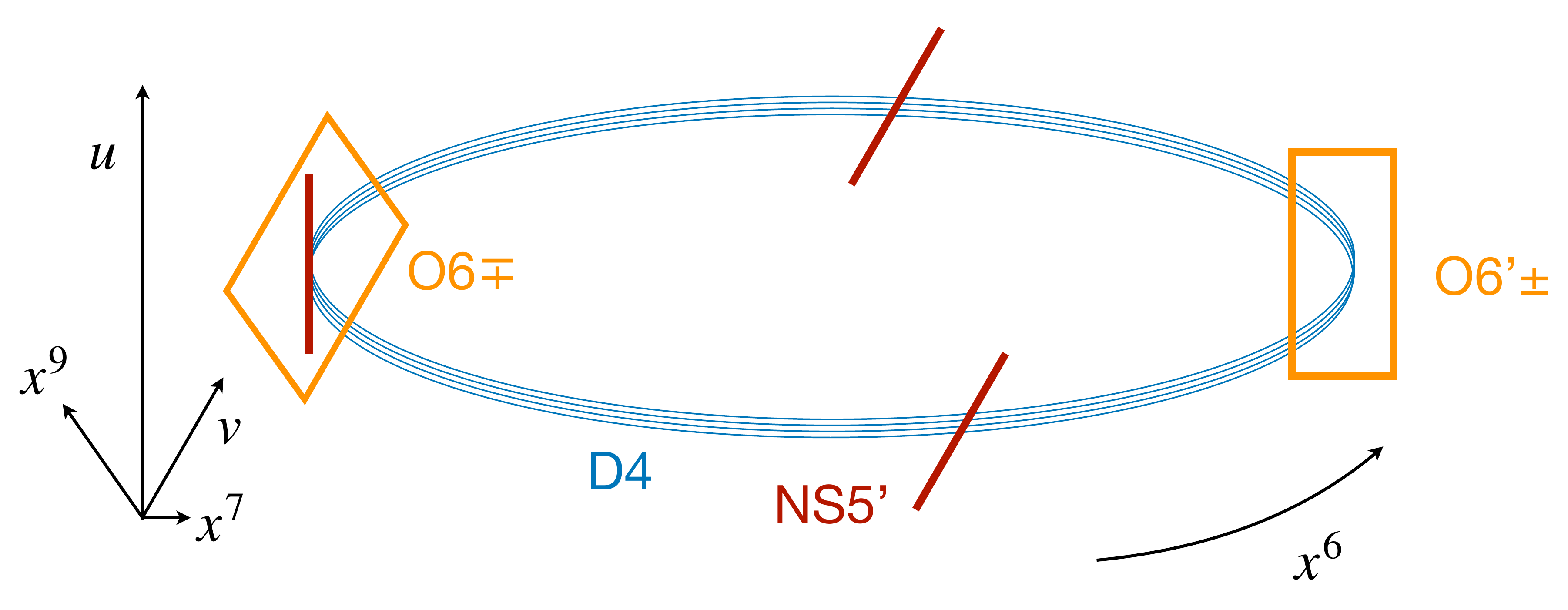}} \quad
	\subfloat
	{\includegraphics[width=.4\textwidth]{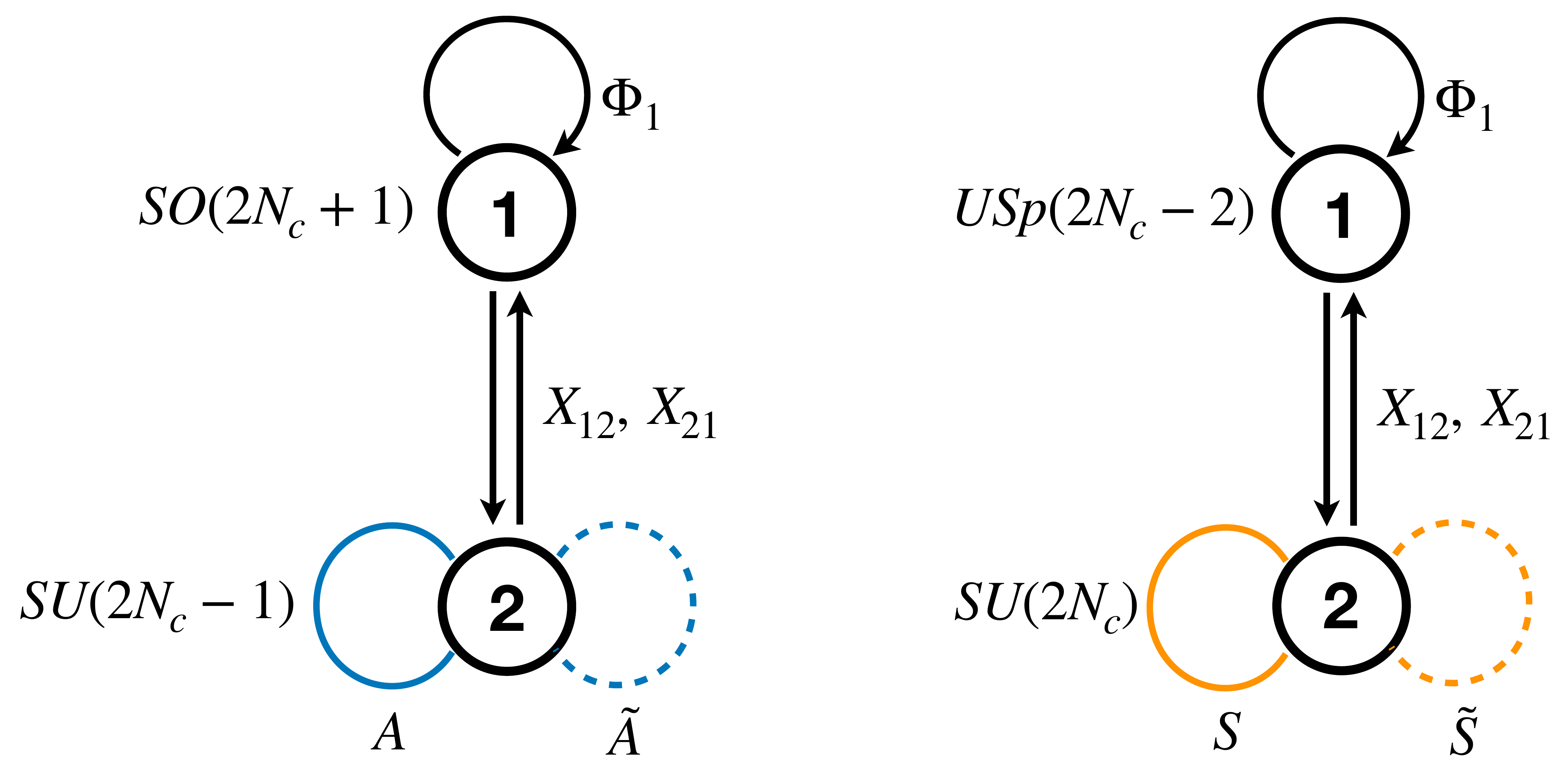}} 
	\caption{Examples of $\mathcal{N}=1$ S-dual quivers with $n_g=1$ obtained from the same $\mathcal{N}=2$ elliptic models by breaking supersymmetry in different ways. All the $\phi_i$ tensor fields have R-charge $1$, therefore we can add the exactly marginal deformation $m_i \phi_i^2$ and integrate them out, yielding the same quiver for the three cases.}
	\label{fig:N1ng1}
\end{figure}

Starting from the elliptic models in figure \ref{fig:N2ng1} we can obtain other S-dual theories by breaking supersymmetry from $\mathcal{N}=2$ to $\mathcal{N}=1$. This can be achieved in different ways, tilting some NS5-branes and/or an O6-plane. We illustrate some examples in figure \ref{fig:N1ng1}.
In the first case supersymmetry is broken by tilting two NS5-branes, i.e. by giving masses to two adjoint fields of the original $\mathcal{N}=2$ theory. Therefore the oriented theory has an $SO$ ($USp$) gauge group with a field $\phi_1$ in the symmetric (antisymmetric) representation and an $SU$ gauge group with two fields in the antisymmetric (symmetric) and conjugate antisymmetric (symmetric) representations.

In the second case supersymmetry is broken by tilting the O6-plane on top of the D4-branes. The resulting theories have an $SO$ ($USp$) gauge group with a field $\phi_1$ in the symmetric (antisymmetric) representation and an $SU$ gauge group  with two fields in the antisymmetric (symmetric) and conjugate antisymmetric (symmetric) representations and a field $\phi_2$ in the adjoint representation.

In the last case both the NS5-branes and the O6-plane are tilted and the resulting theories have an $SO$ ($USp$) gauge group with a field $\phi_1$ in the adjoint representation and an $SU$ gauge group with two fields in the antisymmetric (symmetric) and conjugate antisymmetric (symmetric) representations.

There is a fourth model where the only matter fields are the bifundamentals and the antisymmetric (symmetric) for the unitary group, together with the respective conjugate representation. This is engineered by giving the rightmost O6-plane at an angle in the $(4578)$ directions and considering NS5 and NS5'-branes consistent with the $\zz_2$ quotient induced by the O6-plane. This construction is rather cumbersome and we will not analyze it here because the quiver without any adjoints can be more easily obtained by moving on the conformal manifold of the other cases to a point where all the adjoints are massive and integrating them out.

We will analyze in more detail the third case. The superpotentials of the theories are given by
\begin{equation}
	\begin{split}
		& W_{ \mathcal{A}} = \phi_1 X_{12}X_{21} - X_{12}A \tilde{A} X_{21}\ ,\\
		& W_{ \mathcal{B}} = \phi_1 X_{12}X_{21} - X_{12}S \tilde{S} X_{21}\ .
	\end{split}
	\label{N1ng1:W}
\end{equation}
The central charges are
\begin{equation}
	a_{ \mathcal{A}} = c_{ \mathcal{A}} =  a_{ \mathcal{B}} = c_{ \mathcal{B}} = \frac{81}{128}N_c(2N_c-1)\ .
	\label{N1ng1:ac}
\end{equation}
The matter content and the charges of both the theories are listed in table \ref{N1ng1:tab}.
\begin{table}
	\renewcommand{\arraystretch}{1.5}
	\[
	\begin{array}{lccccc}
		\toprule
		\ & \quad SO(2N_c+1) \quad & \quad SU(2N_c-1) \quad & \quad U(1)_R \quad & \quad U(1)_B \quad & \quad U(1)_2 \quad \\
		\midrule
		\phi_1 & \tiny{\yng(2)} & \mathbf{1} & 1 & 0 & 0 \\
		X_{12} & \tiny{\yng(1)} & \overline{\tiny{\yng(1)}} & 1/2 & 1 & 2N_c-2 \\
		X_{21} & \tiny{\yng(1)} & \tiny{\yng(1)} & 1/2 & -1 & -(2N_c-2) \\
		A & \mathbf{1} & \tiny{\yng(1,1)} & 1/2 & 1 & -2 \\
		\tilde{A} & \mathbf{1} & \overline{\tiny{\yng(1,1)}} & 1/2 & 1 & 2 \\
		\toprule
		\ & \quad USp(2N_c-2) \quad & \quad SU(2N_c) \quad & \quad U(1)_R \quad & \quad U(1)_B \quad & \quad U(1)_2 \quad \\
		\midrule
		\phi_1 &  \tiny{\yng(1,1)} & \mathbf{1} & 1 & 0 & 0 \\
		X_{12} & \tiny{\yng(1)} & \overline{\tiny{\yng(1)}} & 1/2 & 1 & 2N_c-1 \\
		X_{21} & \tiny{\yng(1)} & \tiny{\yng(1)} & 1/2 & -1 & -(2N_c-1) \\
		S & \mathbf{1} & \tiny{\yng(2)} & 1/2 & 1 & 0 \\
		\tilde{S} & \mathbf{1} & \overline{\tiny{\yng(2)}} & 1/2 & -1 & 0 \\
		\bottomrule
	\end{array}
	\]
	\caption{Matter content and charges of $\mathcal{N}=1$ S-dual quivers with $n_g=1$.}  
	\label{N1ng1:tab}
\end{table}
Let us call $b$ the fugacity of $U(1)_B$ and $t$ that of $U(1)_2$; then in the smallest case of $N_c =2$ the superconformal index for both theories reads 
\begin{align}
\mathcal{I} =&\ 1+2 (pq)^{1/2} + (p q)^{1/2}(p+q) + pq \left(b^4+\frac{1}{b^4}+4\right) + \nonumber \\ 
& +(p q)^{3/4} (p+q) \left( -b^3 t^6-\frac{1}{b^3 t^6}-b t^6-\frac{1}{b t^6} \right) + (p q)^{1/2} \left(p^2+q^2\right) + \nonumber \\
&+ p q (p + q) \left( b^4-b^2-\frac{1}{b^2}+\frac{1}{b^4}+2\right) + \nonumber \\ 
&+(p q)^{3/2} \left(8+b^6 t^{12}+b^2 t^{12}+\frac{1}{b^2 t^{12}}+\frac{1}{b^6 t^{12}}+b^4+2 b^2+\frac{2}{b^2}+\frac{1}{b^4}\right) + \ldots\ ;
\end{align}
for $N_c=3$ it reads
\begin{align}
\mathcal{I} = &\ 1+2 (pq)^{1/2} + (p q)^{1/2}(p+q) + 5 pq + (p q)^{3/4} (p+q) \left( -b t^{10}-\frac{1}{b t^{10}} \right) + \nonumber \\  
& +(p q)^{1/2} (p^2+q^2) + 2p q (p + q) +(p q)^{3/2} \left(11 +b^6 +\frac{1}{b^6} + b^2 t^{20} +\frac{1}{b^2 t^{20}}\right) + \ldots\ ,
\end{align}
whereas for $N_c=4$ we were able to compute it only up to the first nontrivial order,
\begin{equation}
\mathcal{I} = 1+2(pq)^{1/2}+ \ldots \ ,
\end{equation}
and it matches across the dual theories.

\subsection{\texorpdfstring{$\mathcal{N}=1$ $n_g=2$}{N=1 ng=2}}
\begin{figure}[t]
	\centering
	\includegraphics[width=0.5\textwidth]{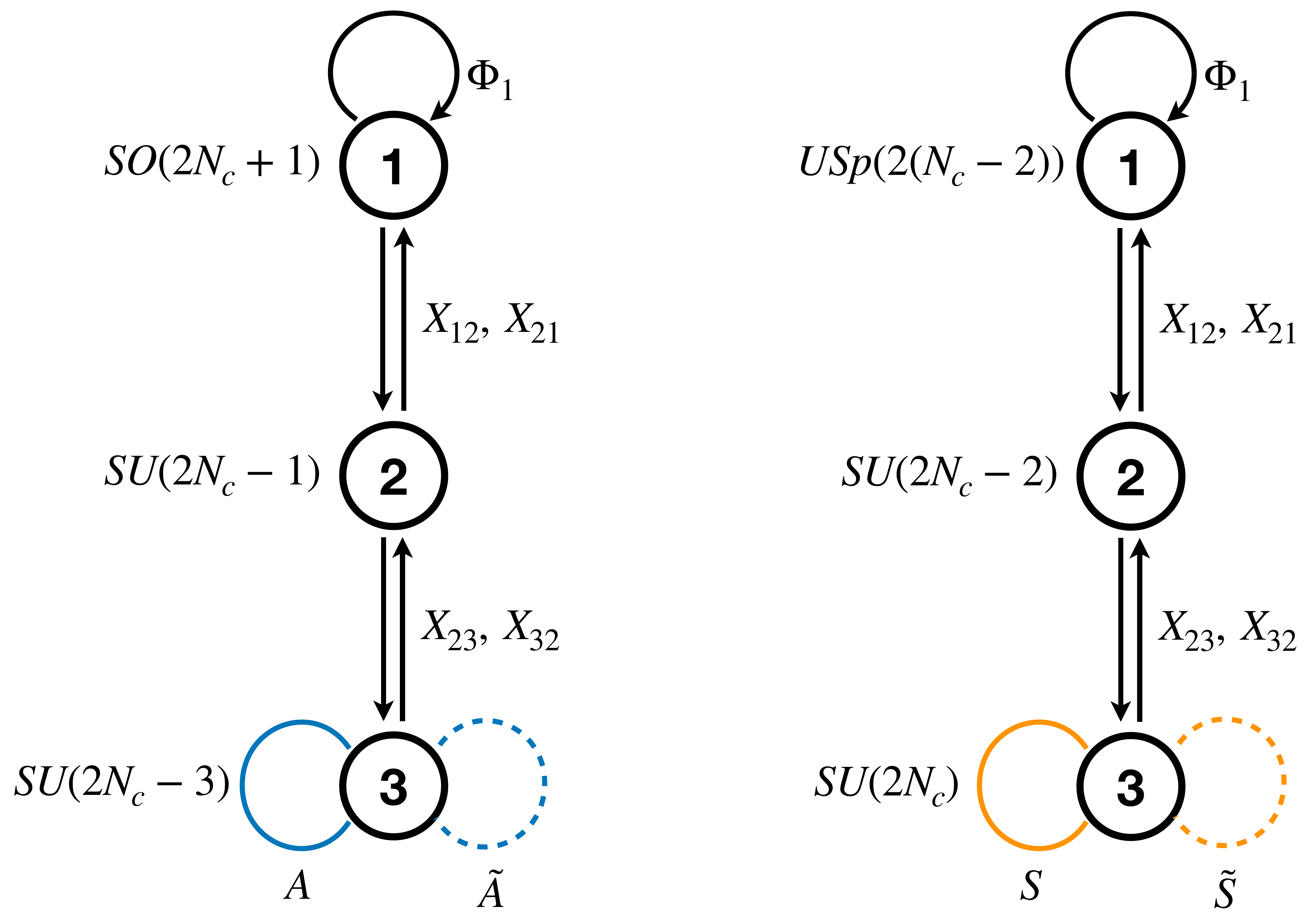}
	\caption{$\mathcal{N}=1$ S-dual quiver diagrams with $n_g=2$.}
	\label{fig:N1ng2}
\end{figure}

As in the $\mathcal{N}=2$ case, adding an $SU$ node to the quivers in figure \ref{fig:N1ng1} and choosing the ranks of the gauge groups in order make the $\beta$ functions vanish, we obtain other $\mathcal{N}=1$ S-dual theories. An example is given in figure \ref{fig:N1ng2}.
The superpotentials of these theories are
\begin{equation}
	\begin{split}
		& W_{ \mathcal{A}} = \phi_1 X_{12}X_{21} - X_{12}X_{23}X_{32}X_{21} + X_{23} A \tilde{A} X_{32}\ ,\\
		& W_{ \mathcal{B}} =  \phi_1 X_{12}X_{21} - X_{12}X_{23}X_{32}X_{21} + X_{23} S \tilde{S} X_{32}\ .
	\end{split}
	\label{N1ng2:W}
\end{equation}
Their matter content and charges are summarized in table \ref{N1ng2:tab}, while their central charges are:
\begin{equation}
	a_{\mathcal{A}} = c_{ \mathcal{A}} = a_{ \mathcal{B}} = c_{ \mathcal{B}} = \frac{27}{128} (8 + 5 N_c (2 N_c-3))\ .
	\label{N1ng2:ac}
\end{equation}
\begin{table}
	\renewcommand{\arraystretch}{1.5}
	\centering
	\begin{adjustbox}{width=\columnwidth,center}
	\renewcommand{\arraystretch}{1.5}
	\begin{tabular}{lccccccc}
		\toprule
		 & $SO(2N_c+1)$ &  $SU(2N_c-1)$  &  $SU(2N_c-3)$ &   $U(1)_R$  & $U(1)_B$ & $U(1)_2$  &  $U(1)_3$ \\
		\midrule
		$\phi_1$ & $\tiny{\yng(2)}$ & $\mathbf{1}$ & $\mathbf{1}$ & $1$ & $0$ & $0$ & $0$ \\
		$X_{12}$ & $\tiny{\yng(1)}$ & $\overline{\tiny{\yng(1)}}$ & $\mathbf{1}$ & $1/2$ & 1 & $2N_c-2$ & $0$ \\
		$X_{21}$ & $\tiny{\yng(1)}$ & $\tiny{\yng(1)}$ & $\mathbf{1}$ & $1/2$ & $-1$ & $-(2N_c-2)$ & $0$ \\
		$X_{23}$ & $\mathbf{1}$ & $\tiny{\yng(1)}$ & $\overline{\tiny{\yng(1)}}$ & $1/2$ & $0$ & $-(2N_c-2)$ & $2N_c-4$ \\
		$X_{32}$ & $\mathbf{1}$ & $\overline{\tiny{\yng(1)}}$ & $\tiny{\yng(1)}$ & $1/2$ & $-1$ & $2N_c-2$ & $-(2N_c-4)$ \\
		$A$ & $\mathbf{1}$ & $\mathbf{1}$ & $\tiny{\yng(1,1)}$ & $1/2$ & $1$ & $0$ & $-2$ \\
		$\tilde{A}$ & $\mathbf{1}$ & $\mathbf{1}$ & $\overline{\tiny{\yng(1,1)}}$ & $1/2$ & $1$ & $0$ & $2$ \\
		\toprule
		 & $USp(2N_c-4)$ &  $SU(2N_c-2)$ &  $SU(2N_c)$ & $U(1)_R$ & $U(1)_B$  &  $U(1)_2$  &  $U(1)_3$ \\
		\midrule
		$\phi_1$ & $\tiny{\yng(1,1)}$  & $\mathbf{1}$ & $\mathbf{1}$ & 1 & 0 & 0 & 0 \\
		$X_{12}$ & $\tiny{\yng(1)}$ & $\overline{\tiny{\yng(1)}}$ & $\mathbf{1}$ & $1/2$ & 1 & $-(2N_c-1) $& $2N_c-3$ \\
		$X_{21}$ & $\tiny{\yng(1)}$ & $\tiny{\yng(1)}$ &$ \mathbf{1} $& $1/2$ & $-1$ & $2N_c-1$ & $-(2N_c-3)$ \\
		$X_{23}$ & $\mathbf{1}$ & $\tiny{\yng(1)}$ &$ {\tiny{\yng(1)}}$ & $1/2$ & 1 & $2N_c-1$ & 0 \\
		$X_{32}$ & $\mathbf{1}$ & $\overline{\tiny{\yng(1)}}$ & $\tiny{\yng(1)} $& $1/2$  & $-1$ & $-(2N_c-1) $& 0 \\
		$S$ & $\mathbf{1}$ & $\mathbf{1}$ &$ \tiny{\yng(2)}$ & $1/2$ & 1 & 0 & 0 \\
		$\tilde{S}$ & $\mathbf{1}$ & $\mathbf{1}$ & $\overline{\tiny{\yng(2)}}$ & $1/2$ & $-1$ & 0 & 0 \\ 
		\bottomrule
	\end{tabular}
\end{adjustbox}
	\caption{Matter content and charges of $\mathcal{N}=1$ S-dual quivers with $n_g=2$.}  
	\label{N1ng2:tab}
      \end{table}%
Let us call $b$ the fugacity of $U(1)_B$, $t$ that of $U(1)_2$, $v$ that of $U(1)_3$; then in the smallest case of $N_c=3$ the superconformal index of both theories reads
\begin{align}
\mathcal{I} =&\ 1+3(pq)^{1/2}+ (p q)^{1/2}(p+q) + 8pq + (p q)^{1/2} (p^2+q^2) + 3 p q (p+q) + \nonumber \\ 
& + (p q)^{3/2}  \left(19+b^6+\frac{1}{b^6}\right) +\ldots \ ,
\end{align}
whereas for $N_c=4$ we are able to compute it only up to the first nontrivial order,
\begin{equation}
\mathcal{I} = 1+3(pq)^{1/2} + \ldots\ ,
\end{equation}
and it matches across the dual theories.

\subsection{\texorpdfstring{The $L^{pqp}$ family}{The L^pqp family}}

We conclude our analysis by discussing the possibility of having fields with R-charge $1$ in toric necklace
quivers without orientifold projections and fractional branes.
The models correspond to $N_c$ D4-branes on a circle with $p+q$ NS and $ p$ NS' fivebranes intersecting 
them at different positions along the compact direction.
These models correspond to the $L^{pqp}$ subfamily of the $L^{pqr}$ theories discovered in \cite{Benvenuti:2005ja,Butti:2005sw,Franco:2005sm}.
They have $p+q$ gauge groups, each pair of nodes is connected by bifundamentals and
there is a `minimal' phase in which there are $q-p$ nodes with an adjoint field.
Other phases with more adjoints are obtained by acting with toric duality.

\begin{figure}
\begin{center}
\includegraphics[width=15cm]{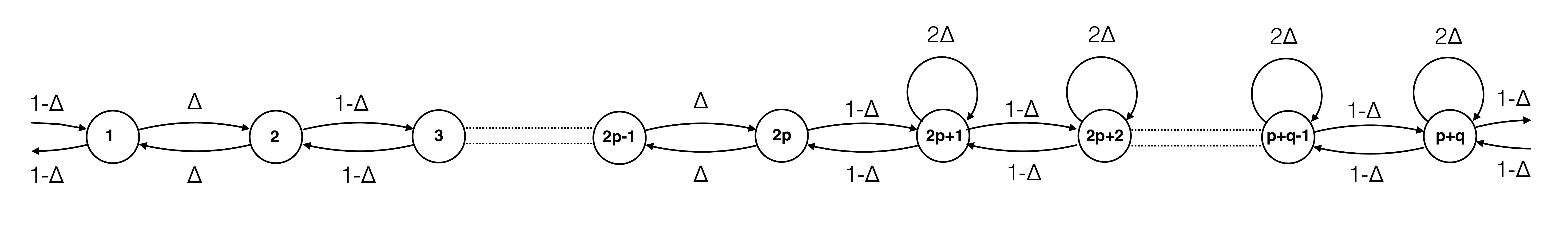}
\caption{Parameterization of the R-charges for the $L^{pqp}$ (necklace) theories. A label over an arrow denotes the R-charge (in terms of $\Delta$) of the corresponding matter field.}
\label{figLpqpDelta}
\end{center}
\end{figure}

When studying the R-charges of the fields one has to maximize the $a$ central charge.
If we order the nodes such that the first $2 p$ nodes appear without any adjoint and we move the 
adjoints on the last $q-p$ nodes, we can parameterize the R-charges of the fields 
as in figure \ref{figLpqpDelta}.
By $a$-maximization we find 
\begin{equation}
\Delta = \frac{2-\frac{p}{q}-\sqrt{\frac{p^2}{q^2}-\frac{p}{q}+1}}{3 \left(1-\frac{p}{q}\right)}\ .
\end{equation}
\begin{figure}
\begin{center}
\includegraphics[width=6cm]{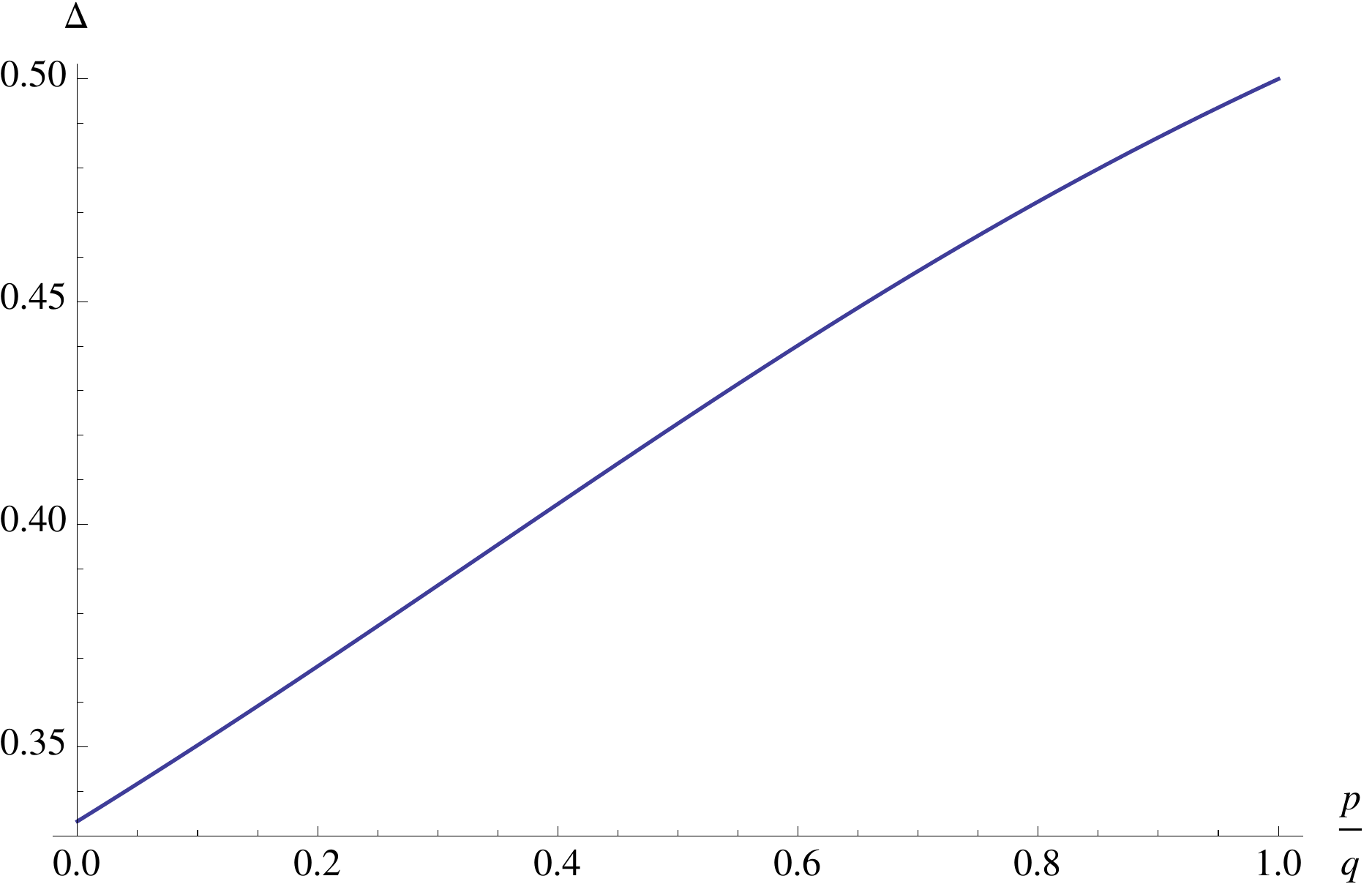}
\caption{Representation of the R-charge $\Delta$ with respect to the ratio ${p}/{q}$ for $L^{pqp}$  models.
Both ${q}$ and ${p}$ are positive integers, with $0<p<q$, i.e. $0<{p}/{q}<1$. The physical R-charges correspond to the ones when the ratio  ${p}/{q}$ is a rational number between $0$ and $1$ in the plot.}
\label{figDelta}
\end{center}
\end{figure}%
We can  study the behavior of this R-charge as a function of the ratio ${p}/{q}$, 
and we observe that the adjoints never have R-charge equal to $1$ in the bulk of the plot in figure \ref{figDelta}.

An interesting case is when $q=p$, with bifundamentals of R-charge
${1}/{2}$. By toric/Seiberg duality one obtains R-charge equal to $1$ for the adjoints.
Furthermore in these cases the dual models can also be obtained  from the $\mathcal{N}=2$ necklaces.
This is done by adding mass deformations for some of the adjoints, and integrating them out.
These models then represent cases where the inherited S-duality can be viewed as Seiberg duality,
generalizing the case of $N_f=2N_c$ SQCD.
This is one of the cases where there is an overlap between Seiberg duality and S-duality. 

On the other hand if we add mass terms on the $\mathcal{N}=2$ necklaces such that the $\mathcal{N}=1$ theories
are not Seiberg-dual (the corresponding $\mathcal{N}=1$ theories are $L^{pqp}$ models with ${q}\neq {p}$), 
 there are no fields with R-charge equal to $1$ and these theories have different central charges
with respect to the $L^{ppp}$ case. Indeed neither S-duality nor Seiberg duality hold between $L^{pqp}$ and $L^{ppp}$
models.

%
%
%
%
\section{Conclusions}
\label{sec:conc}
%
%
%
In this paper we have discussed the notion of inherited S-duality for 4d $\mathcal{N}=1$ quiver gauge theories obtained by an orientifold projection with pairs of O6-planes on Type IIA brane setups with circular D4  and NS5-branes. We have observed that by adding  opportune amounts of  segments of D4-branes between pairs of consecutive NS5-branes in the compact direction, necessary to cancel the anomalous contributions of the O-planes, there are adjoint chiral multiplets with R-charge equal to one. These fields play a central role in our discussion, inducing the notion of S-duality on the conformal manifold, that we dubbed \emph{conformal duality}.

The notion of S-duality for $\mathcal{N}=1$ oriented quivers has been previously discussed in \cite{Garcia-Etxebarria:2012ypj,Bianchi:2013gka,Garcia-Etxebarria:2013tba,Garcia-Etxebarria:2015hua,Garcia-Etxebarria:2016bpb}.  However the models studied in those papers are different from the ones discussed here. In fact, they correspond to toric Calabi--Yau cones over five-dimensional Sasaki--Einstein bases probed by a stack of D3-branes, with the only singularity located at the tip of the cone. Here we have studied cases with extra orbifold singularities. In such cases quivers with a nonchiral-like field content are ubiquitous. Indeed they give rise to the massive deformations that resolve the singularity. The nonchiral field content allows for the presence of adjoint fields as well. These fields, in the cases discussed here, have R-charge equal to one, and they are responsible for the conformal duality.
It would be interesting to translate this analysis to the holographic dual setup, in order to understand the geometric origin of this conformal duality.
Here we restricted our analysis to a subset of models, namely the $L^{pqp}$ family. Nevertheless similar results are expected also for other singular cases, where the matter content is chiral-like, even if there are pairs of nodes connected by a nonchiral-like field content.  This latter type of models has been already considered in \cite{Antinucci:2020yki}, where similar dualities have been obtained. It would be interesting to understand if these dualities correspond to S-dualities as well.

A further line of research consists in studying analogous behaviors for models where extra semi-infinite D6-branes have to be considered on top of the NS5-branes in order to cancel the anomalies. These cases correspond to models with nonabelian flavor symmetries and with fundamental matter fields. 
The simplest case corresponds to family \emph{iv)} with $n_g=0$. The model is not self-dual under S-duality and the strongly coupled dual phases have been discussed in \cite{Chacaltana:2012zy,Chacaltana:2014nya}. There are however also Lagrangian descriptions of such dual phases \cite{Zafrir:2019hps,Etxebarria:2021lmq} with visible  $\mathcal{N}=1$ supersymmetry. It would be interesting to generalize such constructions to larger values of $n_g$.
 
Finally, we observe that the inherited dualities discussed here require the presence of an operator $\mathcal{O}$ with R-charge one and a superpotential deformation $W=\mathcal{O}^2$. This condition corresponds to the \emph{branch flip} in $a$-maximization discussed in \cite{Amariti:2012wc}.
 It may be interesting to have a deeper understanding of this phenomenon connected to the analysis performed here.
 
%
%
%
%
%
%
\section*{Acknowledgments}
%
%
It is a pleasure to thank Massimo Bianchi, I\~naki Garc\'ia-Etxebarria and Kenneth Intriligator for useful comments and for a careful reading of the manuscript.
This work has been supported in part by the Italian Ministero dell'Istruzione, Universit\`a e Ricerca (MIUR), in part by Istituto Nazionale di Fisica Nucleare (INFN) through the ``Gauge Theories, Strings, Supergravity'' (GSS) research project and in part by MIUR-PRIN contract 2017CC72MK-003.
The work of M.F. is supported in part by the European Union's Horizon 2020 research and innovation
programme under the Marie Sk\l odowska-Curie grant agreement No. 754496 - FELLINI.

\bibliographystyle{JHEP}
\bibliography{ref}
\end{document}